\documentclass[a4paper,10pt]{article}
\pdfoutput=1 

\usepackage[T1]{fontenc} 

\usepackage{amsmath}
\usepackage{mathrsfs}
\usepackage{color}
\usepackage{comment}
\usepackage{fancyvrb}
\usepackage{graphicx}
\usepackage{amsmath}
\usepackage{amssymb}
\usepackage{bm}
\usepackage{amsfonts}
\usepackage{times}
\usepackage{epsfig}
\usepackage{dcolumn}
\usepackage{bm}
\usepackage{color}
\usepackage{longtable}
\usepackage{array}
\usepackage{multirow}
\usepackage{verbatim}
\usepackage[utf8]{inputenc}
\usepackage{scalerel}
\usepackage{graphicx}
\usepackage[normalem]{ulem}
\usepackage{subcaption}
\usepackage{lineno}
\usepackage{physics}
\usepackage{wrapfig}
\usepackage{float}

\VerbatimFootnotes

\newcommand{\ee}{e^+ e^-}

\newcommand{\pipi}{\pi^+\pi^-}
\newcommand{\jpsi}{J/\psi}
\newcommand{\pip}{\pi^+}
\newcommand{\pim}{\pi^-}
\newcommand{\piz}{\pi^0}
\newcommand{\KS}{K_{S}}
\newcommand{\KL}{K_{L}}
\newcommand{\Kone}{K_{1}}
\newcommand{\Ktwo}{K_{2}}
\newcommand{\Kp}{K^{+}}
\newcommand{\Km}{K^{-}}
\newcommand{\Kz}{K^{0}}
\newcommand{\Kzbar}{\bar{K}^{0}}

\newcommand{\rt}{\rightarrow}

\newcommand{\BR}{{\cal B}}
\newcommand{\C}{${\mathcal C}$}
\newcommand{\Par}{${\mathcal P}$}

\newcommand{\CPa}{{\mathcal C}{\mathcal P}}
\newcommand{\T}{${\mathcal T}$}

\newcommand{\Ham}{{\mathcal H}}
\newcommand{\Mass}{{\mathcal M}}
\newcommand{\Gmm}{{\boldsymbol \Gamma}}

\newcommand{\Str}{${\mathcal S}$}
\newcommand{\DelS}{\Delta{\mathcal S}}

\newcommand{\eps}{\varepsilon}
\newcommand{\epsp}{\varepsilon^{\prime}}

\newcommand{\Ecm}{E_{\rm cm}}

\newcommand{\smallonehalf}{{\scriptstyle \frac{1}{2}}}

\newcommand{\rootionehalf}{{\textstyle \frac{i}{\sqrt{2}}}}
\newcommand{\smallrootonehalf}{{\scriptstyle \sqrt{\frac{1}{2}}}}

\title{You never have enough $\jpsi$  events}

\author{Stephen Lars Olsen\\
  \\
  Institute for Basic Science,\\
  Daejeon 34126, Republic of Korea\\
  \\
  University of Science and Technology of China,\\
  Hefei 230026, People's Republic of China
}

\makeatletter
\def\@maketitle{%
  \newpage
  \null
  \vskip 2em%
  \begin{center}%
  \let \footnote \thanks
       {\LARGE \@title \par}%
    \vskip 1em%
    {\Large  $-$The case for a $\jpsi$ factory$-$}%
    \vskip 1.5em%
    {\large
      \lineskip .5em%
      \begin{tabular}[t]{c}%
        \@author
      \end{tabular}\par}%
    \vskip 1em%
    {\large \@date}%
  \end{center}%
  \par
  \vskip 1.5em}
\makeatother

\date{\today}

\begin{document}
\maketitle
\begin{abstract}
  In a talk at an IHEP-Beijing symposium celebrating the 50$^{\rm th}$ anniversary of the discovery of the $\jpsi$, I reminisced about
  some of the interesting phenomena that were observed in $\jpsi$ decays in the BES, BESII, and BESIII detectors during the 35
  year long BES experimental program.  The three order of magnitude increase in the $\jpsi$ event samples and the improved
  detector capabilities that occurred during this time  led to a persistent series of interesting  discoveries and
  new insights.  As more $\jpsi$ events were collected, the scientific interest in them increased and the breadth of physics topics that
  are addressed by them expanded. In addition, I speculated on what might happen if, in the next few decades, the magnitude of
  $\jpsi$ data samples and the detector capabilities continued to increase at similar paces. In particular, I emphasize the possibilities of
  searches for \C\Par~violation in strange hyperon decays and testing the \C\Par\T~theorem with strangeness-tagged neutral kaon decays.
\end{abstract}

\section{Introduction}

``Why do we need more $\jpsi$ events?'' and  ``...don't we already have enough?''  are questions that have been frequently asked
at collaboration meetings throughout the history of the BES, BESII and, now, BESIII experiments, usually by charm physics
aficionados while they are clamoring for more data at the $D\bar{D}$ or $D_s\bar{D}_s$ thresholds, or, more recently, by $XYZ$-meson
enthusiasts with ambitions to find yet another multiquark state. In the past, the answer has been primarily been that
the $\jpsi$ is a prolific source of light hadrons that are produced with low backgrounds and well defined quantum numbers that
have supported numerous, and often unique, studies of the spectroscopy of light-quark hadrons and the dynamics of their
production and decays.  Maybe not the most glamorous physics, but still very  interesting and useful.

In the mid-1990s, during the early running of BES and with 8\,million $\jpsi$ events,  we focused on gluon-rich hadronic systems produced
radiative $\jpsi$ decays and mapped out the $J^P$=\,0$^+$ and 2$^+$ $\pipi$, $\piz\piz$ and $\Kp\Km$ resonances in the 1.5-2.0\,GeV mass
region~\cite{ BES:1996nax} as well as the pretty intricate $\eta\pipi$ and $K\bar{K}\pi$ systems in the 1.1-2.0 GeV mass range, that covers
what are now called the $\eta(1405)$/$\eta(1475)$ resonances~\cite{BES:1999axp}. Some of these three-decade-old papers continue to be of
interest and still get occasionally cited~\cite{Nakamura:2023hbt,Achasov:2015uua}.

\begin{figure}[h!]
  \begin{center}
    \includegraphics[width=1.0\textwidth]{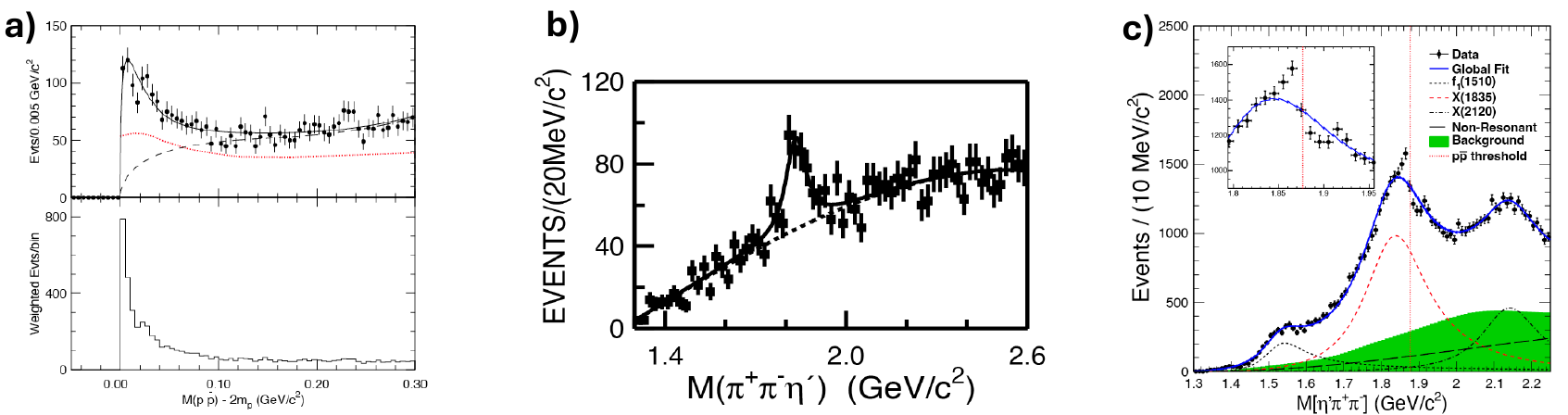}
    \caption{\footnotesize {\bf a)}  The upper panel shows the $M(p\bar{p})$$-$$2m_p$ distribution for $\jpsi$$\rt$$\gamma p\bar{p}$
      in the 58\,M $\jpsi$ event sample~\cite{BES:2003aic}. The fitted curve is described in the text.  The lower panel shows the same
      distribution weighted by the inverse of phase space.
      {\bf b)} The $\pipi\eta^\prime)$ invariant mass distribution for $\jpsi$$\rt$$\gamma\pipi\eta^\prime$ events in the
      58\,M $\jpsi$ event sample~\cite{BES:2005ega}.
      {\bf c)}  The same distribution for 1.09\,B $\jpsi$  events~\cite{BESIII:2016fbr}. The blue curve shows results of a fit
      to the $X(1835)$ peak with a simple Breit-Wigner lineshape.
      }
    \label{fig:X1835}
  \end{center}
\end{figure}

In the BESII era, between 1998 and 2006, 58\,million $\jpsi$ events were collected and, with these, we reported what was the strongest
evidence at the time for the long-sought for $0^+$  $\sigma$  resonance in $\jpsi$$\rt$$\omega\pipi$ decays~\cite{BES:2004mws}
and its strange counterpart, the $\kappa$~in $\jpsi$$\rt$$\bar{K}^{*0}\Kp\pim$ decays~\cite{BES:2005frs}.\footnote{Renamed
       as the $f_0(500)$ and $K^*_0(700)$ by the PDG~\cite{ParticleDataGroup:2024cfk}.}
In addition, we found the curious near-threshold $p\bar{p}$ mass enhancement in $\jpsi$$\rt$$\gamma p\bar{p}$ decays shown in
Fig.~\ref{fig:X1835}a, that, when fitted with a threshold-modified Breit-Wigner function, had a peak mass that was 18\,MeV
{\em below} the 2$m_p$ threshold with a width that was less than 30\,MeV~\cite{BES:2003aic}, and looked very much like a  $p\bar{p}$
bound state. This was followed by the discovery of the $X(1835)$, an $\pipi\eta^\prime$ state with a peak mass in the same
region~\cite{BES:2005ega} (see Fig.~\ref{fig:X1835}b) that was also produced in radiative $\jpsi$ decays. This was interpreted by
some authors as a multigluon decay mode of the subthreshold $p\bar{p}$ state~\cite{Ding:2005gh}.

The BESIII era started operation in 2008, and, with the big boost in statistics provided by 1.09\,B events that
were collected during our first run at the $\jpsi$ peak, we observed strong interference effects on the high-mass side of $X(1835)$
mass peak that occurred exactly at the $M(\pipi\eta^\prime)$=\,2$m_p$ $p\bar{p}$ threshold as shown in Fig.~\ref{fig:X1835}c, where
the fitted curve assumes a simple Breit-Wigner lineshape and the $p\bar{p}$ threshold  is indicated by a dashed vertical line. As more
$\jpsi$ events were accumulated, more intriguing  details became apparent~\cite{BESIII:2016fbr}. At present, this pattern is best described
as the interference of a broad resonance peaked near 1835~MeV and a narrow ($\Gamma$$\sim$10\,MeV) sub-threshold resonance with a peak mass
$\sim$5\,MeV {\em below} 2$m_p$ and with a strong $p\bar{p}$ coupling.

\begin{figure}[h!]
  \begin{center}
    \includegraphics[width=1.0\textwidth]{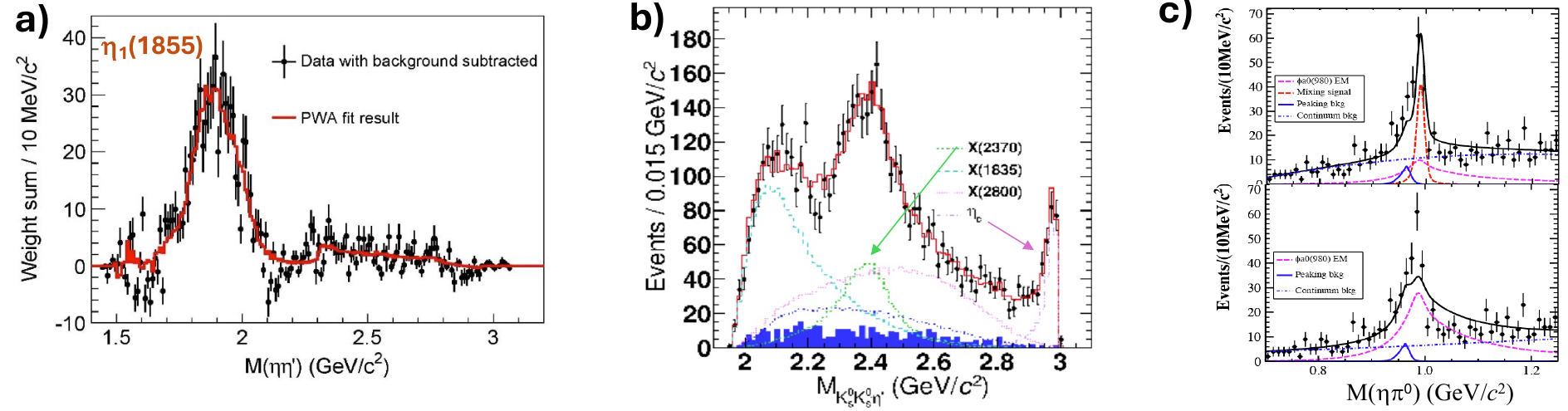}
    \caption{\footnotesize {\bf a)} The  $1^{-+}$~~$\eta\eta^\prime$ partial wave for
      $\jpsi$$\rt$$\gamma\eta\eta^\prime$ events~\cite{BESIII:2022riz}.
      {\bf b)} The $\KS\KS\pipi$ invariant mass distribution for $\jpsi$$\rt$$\gamma\KS\KS\pipi$ events~\cite{BESIII:2023wfi}.  
      {\bf c)} The $\eta\piz$ mass spectrum for $\jpsi$$\rt$$\phi\eta\piz$ events. The top panel shows the results of a fit with
      $\jpsi$$\rt$$\phi f_0(980)$, $f_0$$\rt$$a_0(980)$$\rt\eta\piz$ included and shown as a dashed red curve; the bottom panel shows a fit
      with mixing excluded.~\cite{BESIII:2018ozj}.
      }
    \label{fig:a0-f0-mix}
  \end{center}
\end{figure}

More recently, with 10\,B $\jpsi$ events, BESIII found a strong signal for an $\eta\eta^\prime$ resonance with exotic $J^{PC}$=1$^{-+}$ quantum numbers
in $\jpsi$$\rt$$\gamma\eta\eta^\prime$ decays (with 19$\sigma$ significance!) as shown in Fig.~\ref{fig:a0-f0-mix}a. This state,
called the $\eta_1(1855)$, is a prime candidate for an isoscalar $q\bar{q}$-gluon QCD hybrid state~\cite{BESIII:2022riz}.\footnote{Could
    it be the isoscalar partner of the (also exotic)  $1^{-+}$ $\pi_1(1600)$ candidate hybrid meson~\cite{JPAC:2018zyd}?}
In addition, with $\jpsi$$\rt$$\gamma\Kz\Kzbar\eta^\prime$ decay events in the  same 10\,B $\jpsi$ event sample, BESIII found the $X(2370)$, a
$\Kz\Kzbar\eta^\prime$ resonance with $J^{PC}$=\,0$^{-+}$ (see Fig.~\ref{fig:a0-f0-mix}b), with a mass, width and decay properties
that closely match theoretical predictions  for the lightest pseudoscalar glueball~\cite{BESIII:2023wfi}.

In addition, with billions of  $\jpsi$ events and decays like $\jpsi$$\rt$$\phi f_0(980)$ ($\BR $=\,0.03\%),
$\gamma\eta$ ($\BR $=\,0.1\%),\,\&\, $\gamma\eta^\prime$, ($\BR $=\,0.5\%), BESIII has functioned as a multiple light meson factory with
multimillion-event low-background  samples of tagged $\eta$, $\eta^\prime$\,\&\,$f_0(980)$ decays that have been used to make precision
measurements and support studies of a variety of rare processes (see, {\em e.g.},~\cite{Fang:2021hyq}).

For example, the $\jpsi$$\rt$$\phi f_0(980)$ events  provided a large sample of tagged $f_0(980)$ mesons that were used to measure the
$f_0$$\rt$$a_0$$\rt$$\eta\piz$ mixing process with a significance of 7.4$\sigma$\, as shown in Fig.~\ref{fig:a0-f0-mix}c,~\cite{BESIII:2018ozj}. 
In the same analysis $\psi^\prime$$\rt$$\gamma\chi_{c1}$$\rt$$\gamma\piz a_0(980)$ decays were used to tag $a_0$ mesons that were used to measure the
 $a_0$$\rt$$f_0$$\rt$$\pipi$ reverse process  with 5.5$\sigma$ significance.  These measurement were the first to establish the existence
of $a_0$-$f_0$ mixing, which had been predicted some 40 years earlier~\cite{Achasov:1979xc} but had never been seen. In addition, the
strengths of the  measured $f_0$$\rt$$a_0$ and $a_0$$\rt$$f_0$ couplings strongly supported the four-quark structure of the
$f_0(890)$-$a_0(980)$ nonet of scalar mesons~\cite{Achasov:2019vcs}, something that has been a subject of numerous discussions in the light
hadron community ever since the possibility was first suggested by Bob Jaffe in 1976~\cite{Jaffe:1976ig}.

Six years ago, I reviewed some of the above results   in a talk that I gave at the IHEP symposium celebrating the 30$^{\rm th}$
anniversary of the start of the BES experimental program, and concluded my remarks with the comment:

\begin{figure}[h!]
  \begin{center}
    \includegraphics[width=0.6\textwidth]{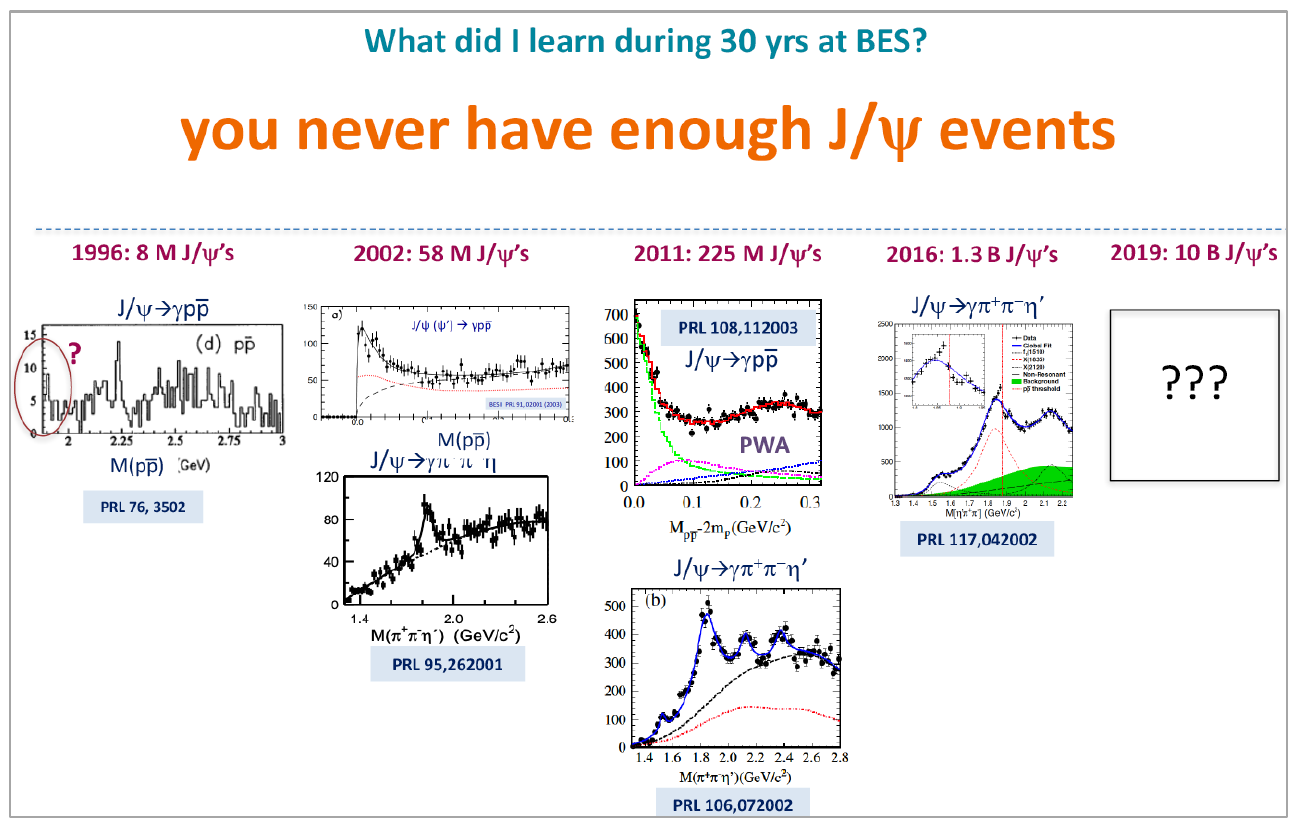}
    \caption{\footnotesize My final slide at the BES 30$^{\rm th}$ anniversary symposium (September 2019).
      }
    \label{fig:enough}
  \end{center}
\end{figure}

\noindent
This comment seemed to  strike a resonant chord in the psyche of my BES colleagues because, in spite of the numerous (what I thought were)
sage observations that I  made during my many years of participation in BES, this is the only one that seems to be remembered and repeated,
and, I suppose, is the reason I was given the honor to speak at this auspicious meeting.

During the BES experimental program, we enjoyed a three order of magnitude increase in the number of $\jpsi$ events we had to work
with, and with each major increase we were rewarded with new insights and discoveries. There was never any sign of saturation; as more $\jpsi$
became available, we always found more things to do with them.  Given this history, it is interesting to
speculate about what might happen if at some post-BESIII, but near-future facility, the next generation of researchers also got to enjoy similar
multiple order-of-magnitude boosts in $\jpsi$ event samples and an even more capable detector.  An obvious extrapolation from the past is that
more light-hadron states would be found, the spectroscopies of QCD glueballs and hybrid mesons would  be established and to a large extent mapped
out, decays of the $\eta$, $\eta^\prime$ and $f_0(980)/a_0(980)$ mesons would be studied with even higher precision and even rarer process would be
found, and, in an era where Lattice QCD and AI  will play an increasingly important role in our science, the large number of $\jpsi$ events would
enable  numerous precision measurements of various quantities that could help calibrate and validate these computations. I won't discuss these today
and, instead, point toward two qualitatively different research opportunities that are close to my heart and that multi-trillion $\jpsi$ event samples
in a more specialized detector would support: searches for \C\Par~violation in strange hyperon decays and tests of the \C\Par\T~theorem with
strangeness-tagged $\Kz$ and $\Kzbar$ decays.

\section{Why \C\Par~ violations in hyperon decays?}

The only source of \C\Par~violation in the SM is the Kobayashi-Maskawa phase in the CKM quark-flavor mixing matrix, and this
has provided precise quantitative explanations for virtually all experimentally observed \C\Par-violating phenomena, including
\C\Par-symmetry violations in the decays of strange and bottom-flavored hadrons. However the KM \C\Par-violating mechanism cannot
explain how the current all-matter universe evolved from what had to have been a matter-antimatter symmetric condition that existed
shortly after the Big Bang. KM-model-based calculations fail to account for the  baryon asymmetry of the current universe---commonly
referred to as the BAU---by some ten orders of magnitude~\cite{Rubakov:1996vz}. Thus, there must have been other, non-SM source
(or sources) of \C\Par~violation in Nature.

Searches for anomalous \C\Par~violations in the $b$-and $c$-quark sectors are the major motivations for the LHCb experiment~\cite{LHCb:2008vvz}
at CERN Large Hadron Collider and the Belle~II experiment~\cite{Aushev:2010bq} at KEK. Although many \C\Par-violations have been established in
the $b$-quark sector, so far at least, these can all be explained by the KM mechanism~\cite{Gershon:2024pdg}, including the recently seen
\C\Par~violation in $\Lambda_B$$\rt$$p\Km\pipi$ decay~\cite{LHCb:2025ray} (see ref.~\cite{He:2025msg}).  For a long time, theorists assured
experimentalists that \C\Par~violating asymmetries in the $c$-quark sector larger than $\sim$0.01\% could not be explained by the KM-mechanism
and would be clear evidence for new, beyond-the-SM physics~\cite{Grossman:2006jg}. This confidence lasted until a few days after the LHCb group
announced evidence for a huge, nonzero \C\Par-violating asymmetry in $D$$\rt$$\Kp\Km$ and $\pipi$ decay in 2011~\cite{LHCb:2011osy}, which was
followed by a 5$\sigma$ discovery of a no-longer huge---but still large---\C\Par~violation in 2019~\cite{LHCb:2019hro}:
\begin{equation}
\nonumber
\textstyle{\frac{\BR(D\rt\Kp\Km)-\BR(\bar{D}\rt\Kp\Km)}{\BR(D\rt\Kp\Km)+\BR(\bar{D}\rt\Kp\Km)}  
-\frac{\BR(D\rt\pipi)-\BR(\bar{D}\rt\pipi)}{\BR(D\rt\pipi)+\BR(\bar{D}\rt\pipi)}}
=-(0.15\pm 0.03)\%.
\end{equation}
Regrettably, instead of being endorsed as new physics discoveries, this \C\Par~violation was attributed to SM processes  enhanced  by
$SU(3)$-violating $\Kp\Km$ and $\pipi$ final-state interactions (see refs.~\cite{Cheng:2012xb} and \cite{Grossman:2012ry}), {\em i.e.}, SM
light-hadron physics in the 1.8-1.9~GeV mass region (that was very familiar to old-timers in the BES collaboration).

Stringent limits on flavor conserving non-SM \C\Par~violations in the ($u,d$) light quark sector have been established by very sensitive searches
for electric dipole moments of the neutron~\cite{Abel:2020pzs} and the electron~\cite{ACME:2018yjb,Roussy:2022cmp}.  And, despite  a barrage of
public statements to the contrary, there is no possible value of the Dirac phase in the PMNS neutrino-mixing matrix that could be measured by
impending long-baseline neutrino experiments~\cite{Abe:2018uyc,DUNE:2020jqi} that could resolve the BAU conundrum~\cite{Davidson:2008bu}.

That leaves us with \C\Par~violations in the $s$-quark sector, where the only direct \C\Par-violation that has been seen so far is the one
characterized by the parameter $\varepsilon^{\prime}$ that was measured as a  small, nonzero effect in $\KL$$\rt$$\pi\pi$ decay by the NA48
experiment at CERN~\cite{NA48:2001bct} and the KTeV experiment at Fermilab~\cite{KTeV:1999kad}. This parameter quantifies the \C\Par-violating
phase difference between interfering isospin\,=\,2 and isospin\,=\,0, parity-violating $S$-wave amplitudes in $K^0(\bar{K}^0)$$\rt$$\pi\pi$
decays, and is suppressed in magnitude by a nominal factor of 1/22.5 by the weak-interaction's $\Delta I$=1/2~rule. The
\C\Par-phase difference\footnote{Unlike
zillions of other phases in physics, a \C\Par~phase has opposite sign for particles and antiparticles.}
between the I\,=\,2 and I\,=\,0 $S$-wave  $K$$\rt$$\pi\pi$ amplitudes that is deduced\footnote{This is
  $\xi^S_{K\rt\pi\pi}$$\approx$$22.5$$\times$$\sqrt{2}\Re(\varepsilon^{\prime}/\varepsilon)$$\times$$\varepsilon$,
  where $|\Im A^{I=2}_{\pi\pi}|$\,=\,$\sqrt{2}|\Re(\varepsilon^{\prime}|/\varepsilon)|$, of which
  $\Re(\varepsilon^{\prime}/\varepsilon)$\,$=$1.7$\times$10$^{-3}$ is what is measured,
  $\varepsilon$\,$=$2.23$\times$10$^{-3}$ is the neutral kaon mass-matrix \C\Par-violating
  parameter, and the 22.5 factor accounts for the larger $\Delta I$=1/2 interfering $A^{I=0}_{\pi\pi}$ amplitude.}
from the measured value of $\varepsilon^{\prime}$
is $\xi^S_{K\rt\pi\pi}$$\approx$$\,0.005^{\circ}$, and is within errors of the SM prediction~\cite{Buras:2020wyv}
$\xi^S_{K\rt\pi\pi}(SM)$\,$\approx$\,$0.004^{\circ}$.

But this only constrains new physics sources of \C\Par~violations in in parity-violating $S$-wave amplitudes. Hyperon decays
involve interfering $S$- and $P$-wave amplitudes.  The measured \C\Par-phase in $K$$\rt$$\pi\pi$ decay and its agreement with
SM expectations has been translated~\cite{Tandean:2002vy,He:1999bv} into stringent limits on contributions from BSM new physics
sources to the $S$-wave amplitudes in hyperon decays that imply that any observed \C\Par~violation 
at the current levels of experimental sensitivity  would correspond to new physics contributions to
the parity-conserving $P$-wave amplitudes.

\subsection{$\boldsymbol{\ee$$\rt$$\jpsi$$\rt$$\Xi\,\bar{\Xi}}$, near-perfect fits for hyperon \C\Par~studies}

\begin{figure}[h!]
  \begin{center}
    \includegraphics[width=0.85\textwidth]{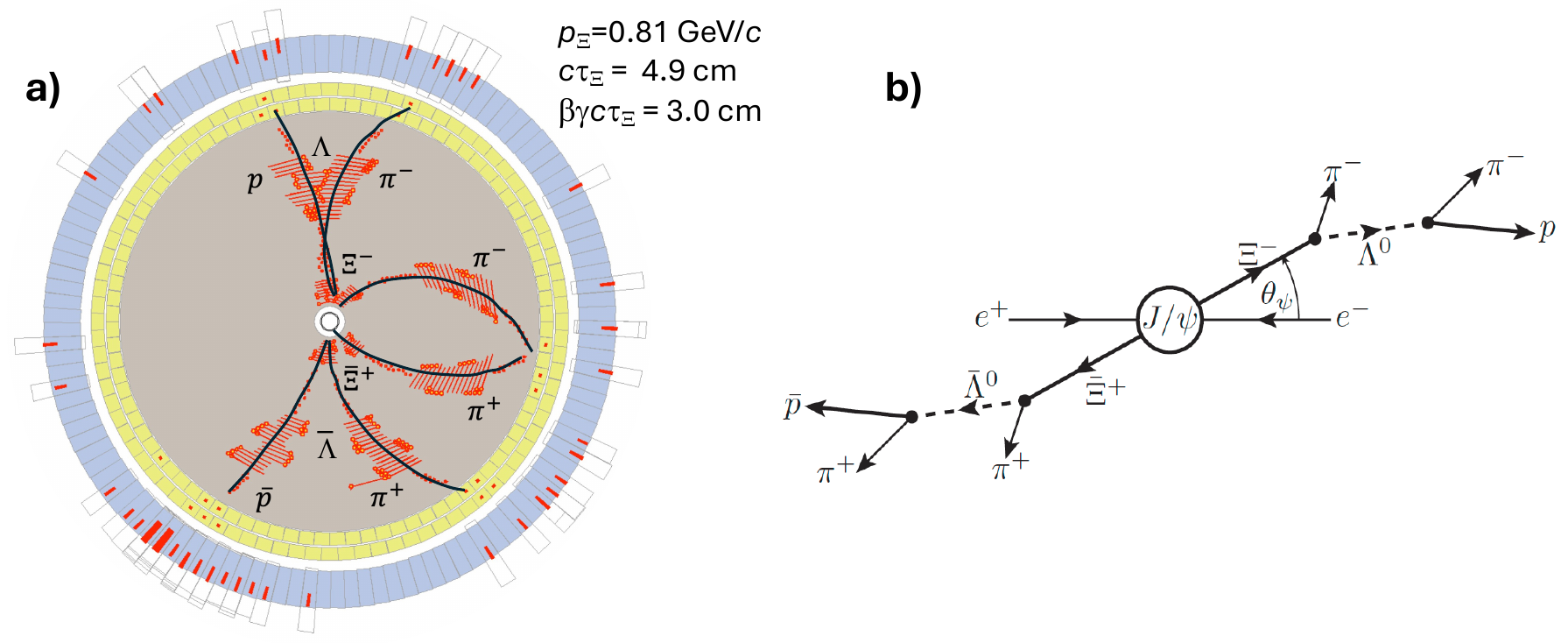}
    \caption{\footnotesize {\bf a)} An $\ee$$\rt$$\jpsi$$\rt$$\Xi^-\bar{\Xi}^+$ event in the BESIII detector.
    {\bf b)} A road map for these events. 
    }
    \label{fig:Jpsi-XiXib}
  \end{center}
\end{figure}

Figures~\ref{fig:Jpsi-XiXib}a shows what an $\ee$$\rt$$\jpsi$$\rt$$\Xi^-\bar{\Xi}^+$, where $\Xi^-$$\rt$$\Lambda\pim$, $\Lambda$$\rt$$p\pim$,
and $\bar{\Xi}^+$$\rt$$\bar{\Lambda}\pip$, $\bar{\Lambda}$$\rt$$\bar{p}\pip$ event looks like in the BESIII detector;
Fig,~\ref{fig:Jpsi-XiXib}b shows a roadmap of the underlying decay chains. These events, along with their $\Xi^0\bar{\Xi}^0$ isospin
counterparts, are nearly perfect reactions for studying hyperon \C\Par~violations for the following reasons:

\vspace{1.5mm}
\noindent
$i$)~the $\Xi$ and $\bar{\Xi}$ are in a mixture of $J^P$=$1^-$ quantum states with $J_z$=$\pm1$ but not $J_z$=\,0;

\vspace{1.5mm}
\noindent
$ii$)~\Par~conservation in their production makes the $\Xi$\,\&\,$\bar{\Xi}$ polarizations
exactly equal;

\vspace{1.5mm}
\noindent
$iii$)~\Par~violation in $\Lambda$\,\&\,$\bar{\Lambda}$ decay provide event-by-event $\Lambda$\,\&\,$\bar{\Lambda}$ spin measurements;

\vspace{1.5mm}
\noindent
$iv$)~quantum correlations \& $\Lambda$ spin measurements make the effective polarizations=1;

\vspace{1.5mm}
\noindent
$v$)~the $\Xi$ and $\bar{\Xi}$ are back-to-back in a symmetric detector with equal momenta;

\vspace{1.5mm}
\noindent
$vi$)~~$\Xi$ and $\bar{\Xi}$ decay products have identical fiducial volumes and selection requirements;

\vspace{1.5mm}
\noindent
$vii$)~most of the $\Xi$\,\&\,$\bar{\Xi}$ decays occur in a vacuum.\\

The overall process involves five different reactions that are most simply described by considering five different
reference frames. The first is the $\jpsi$ rest frame where, since the spin=1 $\jpsi$ mesons are produced in
$\ee$~annihilation via a single virtual photon, they are spin-aligned along the direction of the $\ee$ beamline
(taken as the $z$ axis). This means that equal numbers of them are produced in the $\ket{J;J_z}$=\,$\ket{1;+1}$ and
$\ket{1;-1}$ angular momentum states, but none with $\ket{J;J_z}$=\,$\ket{1;0}$. The $\jpsi$$\rt$$\Xi\,\bar{\Xi}$ decay
matrix element is the sum of two  helicity amplitudes. The interference between these two
amplitudes and the absence of any initial-state $J_z$=\,0 component produces a $\theta_\psi$-dependent production cross
section for $\Xi$~\&~$\bar{\Xi}$ hyperons that are spin-polarized with polarizations 
$\vec{\mathcal{P}}_{\Xi}$  that are perpendicular to the production plane:~\cite{Perotti:2018wxm}:
  \begin{eqnarray}
   \label{eqn:jpsi-xixibar-alpha}
   \frac{1}{N}\frac{dN}{d\cos\theta_{\psi}}&=&\frac{3}{4\pi}\frac{1+\alpha_\psi \cos^2\theta_{\psi}}
        {3+\alpha_{\psi}}\\
       \label{eqn:jpsi-xixibar-beta}  
         \vec{\mathcal{P}}_{\Xi}&=&\frac{\sqrt{1-\alpha^2_{\psi}}\sin\theta_{\psi}\cos\theta_{\psi}\sin\Delta\Phi}
            {1+\alpha_{\psi}\cos^2\theta_{\psi}},
  \end{eqnarray}
where $\alpha_{\psi}$ is a parity- and \C\Par-conserving parameter that characterizes the relative strengths of the two
helicity amplitudes, $\Delta\Phi$ is their relative phase, and the polarizations of the hyperons are equal,
{\it i.e.}, $\vec{\mathcal{P}}_{\Xi}=\vec{\mathcal{P}}_{\bar{\Xi}}$.

\begin{figure}[h!]
  \begin{center}
    \includegraphics[width=0.95\textwidth]{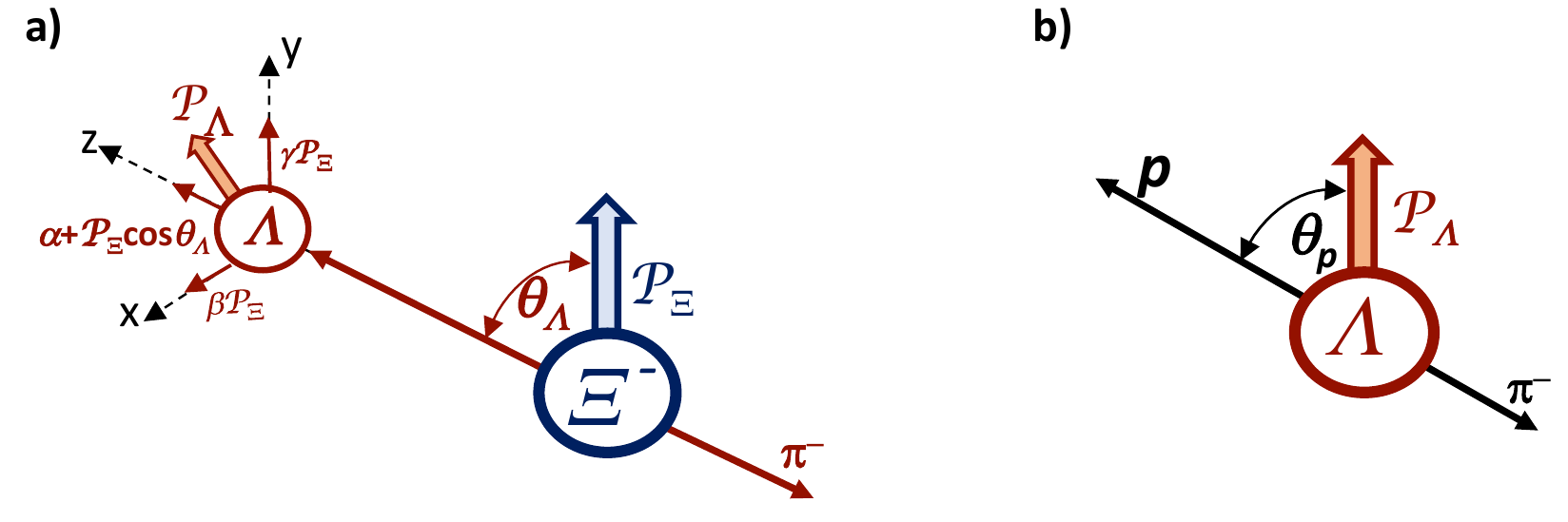}
    \caption{\footnotesize {\bf a)} The decay angle $\theta_{\Lambda}$  and $\Lambda$ polarization vector
      $\mathcal{P}_{\Lambda}$ for polarized $\Xi^-\rt\Lambda\pim$ decays. {\bf b)} The decay angle $\theta_p$
      for polarized $\Lambda\rt p\pim$ decays.}
    \label{fig:Xi_and_Lambda-decays}
  \end{center}
\end{figure}

  The next two reference frames to consider are the $\Xi$ and $\bar{\Xi}$ rest frames. In these frames
  the decay angle $\theta_\Lambda$ ($\theta_{\bar{\Lambda}}$) relative to $\mathcal{P}_{\Xi}$ 
  direction, and the polarization vectors of the $\Lambda$ daughters  $(\vec{\mathcal{P}}_{{\Lambda}})$ and
  $(\vec{\mathcal{P}}_{\bar{\Lambda}})$, are distributed according to~\cite{Lee:1957he}
    \begin{eqnarray}
     &~& \frac{dN}{d\cos\theta_{\Lambda}}\propto 1+\alpha_\Xi\mathcal{P}_{\Xi}\cos\theta_{\Lambda}\\
      \label{eqn:Lambda-pol}
     &~&~~\mathcal{P}_{\Lambda}=\frac{(\alpha_\Xi+\mathcal{P}_{\Xi}\cos\theta_\Lambda)\mathbf{\hat{z}}
      +{\mathcal P}_{\Xi}\beta_\Xi\mathbf{\hat{x}}
      +{\mathcal P}_{\Xi}\gamma_\Xi\mathbf{\hat{y}}}{1+\alpha_{\Xi}\mathcal{P}_{\Xi}\cos\theta_\Lambda},
    \end{eqnarray}
    where the unit vectors $\mathbf{\hat{x},\hat{y},\hat{z}}$ are oriented along the axes indicated in
    Fig.~\ref{fig:Xi_and_Lambda-decays}a  and $\alpha,\beta,\gamma$ are the Lee-Yang decay parameters that are
    defined below.

The last two reference frames are the $\Lambda$ and $\bar{\Lambda}$ rest frames. Here the distribution of
events in $\theta_p$ ($\theta_{\bar{p}}$), the direction of the proton relative to the $\Lambda$ polarization
vector (see Fig.~\ref{fig:Xi_and_Lambda-decays}b), is
\begin{equation}
        \label{eqn:Lambda-ppi}
      \frac{dN}{d\cos\theta_{p}} \propto 1+\alpha_\Lambda\mathcal{P}_{\Lambda}\cos\theta_{p}.
    \end{equation}
 Since the proton and antiproton polarizations are not measured, only the $\alpha_{\Lambda}$ and
 $\alpha_{\bar{\Lambda}}$ parameters can be determined

 The (real) Lee-Yang parameters $\alpha_Y$,~$\beta_Y$~and~$\gamma_Y$  are defined in terms of the (complex) $S$- and $P$-wave decay
 amplitudes $S_Y$ and $P_Y$ as~\cite{Lee:1957he}
\begin{equation}
  \alpha_Y = \frac{2\Re(S^*_YP_Y)}{|S_Y|^2+|P_Y|^2},~~~~\beta_Y = \frac{2\Im(S^*_YP_Y)}{|S_Y|^2+|P_Y|^2},
  ~~~~\gamma_Y = \frac{|S_Y|^2-|P_Y|^2}{|S_Y|^2+|P_Y|^2},
\end{equation}
were $Y$=\,$\Xi$~or~$\Lambda$  and $\alpha^2_Y$+$\beta^2_Y$+$\gamma^2_Y$=1. The $S_Y$ and $P_Y$  decay  amplitudes can have
two distinct types of phases. One type corresponds the $S$- and $P$-wave $\Lambda\pi$
and $p\pi$ final-state strong interaction phase shifts, $\delta^S_{y\pi}~\&~\delta^P_{y\pi}$, where
$Y$$\rt$$y\pi$~($y$=$\Lambda~{\rm or}~p$), and have the {\it same sign} for particles and antiparticles. If \C\Par~is
violated, they will have additional so-called {\it weak}-, or \C\Par-phases
$\xi^{P}_{Y}$ and $\xi^{S}_{Y}$ that have {\it opposite signs} for particles and antiparticles. Thus, $\alpha_Y$
(hyperon) and $\alpha_{\bar{Y}}$ (anti-hyperon) have the form\cite{Donoghue:1985ww}
\begin{eqnarray}
  \alpha_Y&=&
  \frac{2|S_Y||P_Y|\cos\big((\delta^P_{y\pi}-\delta^S_{y\pi})+(\xi^P_{Y}-\xi^S_{Y})\big)}{|S_Y|^2+|P_Y|^2}\\
\nonumber
  \alpha_{\bar{Y}} &=&
  -\frac{2|S_Y||P_Y|\cos\big((\delta^P_{y\pi}-\delta^S_{y\pi})-(\xi^P_{Y}-\xi^S_{Y})\big)}{|S_Y|^2+|P_Y|^2},
\end{eqnarray}
where the overall minus sign in the expression for $\alpha_{\bar{Y}}$ comes from the operation of the parity operator
({\it i.e.} the \Par~in \C\Par) on the $P$-wave amplitude. Thus, assuming $\xi^P_{Y}$$-$$\xi^S_{Y}$ is small,
\begin{equation}
  \label{eqn:ACP-asymmetry}
        {\mathcal A}_{\CPa}^Y\equiv\frac{\alpha_Y+\alpha_{\bar{Y}}}{\alpha_Y-\alpha_{\bar{Y}}}
        \approx -\sin(\delta^P_{y\pi}-\delta^S_{y\pi})(\xi^P_{Y}-\xi^S_{Y}) ,
\end{equation}
and, if \C\Par~is not conserved, $\xi^P_{Y}$$\neq$$\xi^S_{Y}$, and any significant difference between
${\mathcal A}_{\CPa}^Y$ and zero would unambiguous evidence for a \C\Par~violation.

However, it is evident from eqn.~\ref{eqn:ACP-asymmetry} that even if $\xi^P_{Y}$$-$$\xi^S_{Y}$ is nonzero,
${\mathcal A}_{CP}^Y$ would be zero if $\delta^P_{y\pi}$$-$$\delta^S_{y\pi}$=\,0. Although it is unlikely that
these strong-interaction phase-shift differences are exactly zero, they are known to be small---theoretical
estimates are~\cite{Hoferichter:2015hva} $\delta^P_{p\pi}$$-$$\delta^S_{p\pi}$=$-7.3^{\circ}$\,$\pm$1.2$^{\circ}$
and~\cite{Huang:2017bmx} $\delta^P_{\Lambda\pi}$$-$$\delta^S_{\Lambda\pi}$=8.8$^{\circ}$\,$\pm$2.2$^{\circ}$---and these will
suppress the values of ${\mathcal A}_{CP}^Y$ by an order of magnitude. This is not the case for
measurements of the parameter $\beta_{\Xi}$ that can be determined from the measurement of
${\mathcal P}_{\Lambda}$ in $\Xi$$\rt$$\Lambda\pi$ decays. Here, we apply the above-described calculation to
a different asymmetry parameter ${\mathcal B}_{CP}^{\Xi}$, where 
\begin{equation}
  \label{eqn:BCP-asymmetry}
        {\mathcal B}_{CP}^{\Xi}\equiv\frac{\beta_{\Xi}+\beta_{\bar{\Xi}}}{\alpha_{\Xi}-\alpha_{\bar{\Xi}}}
    = \tan(\xi^P_{\Xi}-\xi^S_{\Xi})\approx \xi^P_{\Xi}-\xi^S_{\Xi}.
\end{equation}
In this case the measured asymmetry doesn't suffer any dilution by small strong-interaction
phase-shifts.  In $\Xi\,\bar{\Xi}$ events where all the tracks in the decay chain are detected and reconstructed,
$\beta_\Xi$, $\beta_{\bar{\Xi}}$ and $\mathcal{B}_{CP}^{\Xi}$ can be measured.  This is only possible with $\Xi$ decays.
The $\Lambda$ and $\Sigma$ hyperons produced via the $\jpsi$$\rt$$\Lambda\bar{\Lambda}$ or $\Sigma\bar{\Sigma}$ reactions
decay to nucleons and antinucleons, and precision polarization measurements of nucleons are difficult and impossible for
antinucleons.

Since the $\beta_{\Xi}$ and $\gamma_{\Xi}$ parameters are not independent, BESIII reports
measurements of $\phi_{\Xi}$ (and $\phi_{\bar{\Xi}}$) that were defined by Lee and Yang to be~\cite{Lee:1957he}
\begin{equation}
  \beta_{\Xi}=\sqrt{1-\alpha^2_{\Xi}}\sin\phi_{\Xi}~~{\rm and}
  ~~\gamma_{\Xi}=\sqrt{1-\alpha^2_{\Xi}}\cos\phi_{\Xi}.
\end{equation}
In terms of  $\Delta\phi_{CP}^{\Xi}$\,$\equiv$\,${\textstyle \frac{1}{2}} (\phi_{\Xi}$+$\phi_{\bar{\Xi}})$ and
$\langle\phi_{CP}^{\Xi}\rangle$\,$\equiv$\,${\textstyle \frac{1}{2}} (\phi_{\Xi}$$-$$\phi_{\bar{\Xi}})$:
\begin{eqnarray}
  \label{eqn:Deltaphi_CP}~
  \xi^P_{\Xi}-\xi^S_{\Xi}&=&\frac{\sqrt{1-\alpha^2_{\Xi}}}{\alpha_{\Xi}}\Delta\phi_{CP}^{\Xi} = -2.43\Delta\phi_{CP}^{\Xi}\\
  \nonumber
  \delta^{P}_{\Lambda\pi}-\delta^{S}_{\Lambda\pi}&=&
        \frac{\sqrt{1-\alpha^2_{\Xi}}}{\alpha_{\Xi}}\langle\phi_{CP}^{\Xi}\rangle = -2.43\langle\phi_{CP}^{\Xi}\rangle,
\end{eqnarray}
where, in the far r.h.s.~relation $\alpha_\Xi$ was set at $-0.38$\,.

\subsection{The BESIII $\boldsymbol{\Xi^-\bar{\Xi}}$ and $\boldsymbol{\Xi^0\bar{\Xi}^0}$ event samples}

In the BESIII measurement, the $\Xi$ hyperons occur in the  reaction chains shown in Fig.\ref{fig:Jpsi-XiXib}.
Here the $\jpsi$ mesons are produced (nearly) at rest in the laboratory by counter-circulating $e^+$ and
$e^-$ beams in the BEPCII collider and the $\Xi$~\&~$\bar{\Xi}$ are produced back-to-back, in a spin-correlated
quantum-entangled state that persists until one of them decays, which usually occurs within a few~centimeters of the
production point and in a vacuum. In the selected events, both of the $\Xi$ hyperons decay to a pion and a $\Lambda$
hyperon that, in turn, decays into a charged pion and a proton also within about a few centimeters of the parent $\Xi$ decay
points. The events show up in the detector as $p\pim$ and $\bar{p}\pip$ decay products of nearly back-to-back $\Lambda$ and
$\bar{\Lambda}$ hyperons that are accompanied by two low momentum $\pim$ and $\pip$ tracks in the case of $\Xi^-\bar{\Xi}^+$
events, and one low momentum $\pim$ and $\pip$ track  and two low-energy $\piz$$\rt$$\gamma\gamma$ decays for
$\Xi^0\bar{\Xi}^0$ events.  The detected events have a very clear topology, with highly constrained kinematics and no particle
identification requirements: the highest momentum positive and negative tracks are always the proton and antiproton. The
BESIII 10\,B $\jpsi$-event data sample contained 576\,k fully reconstructed $\Xi^-\bar{\Xi}^+$ events with less than 0.5\%
background, and 321\,k $\Xi^0\bar{\Xi}^0$ events
with less than  2\% percent background.

\subsection{Analysis and results}

\begin{figure}[h!]
   \begin{center}
    \includegraphics[width=0.65\textwidth]{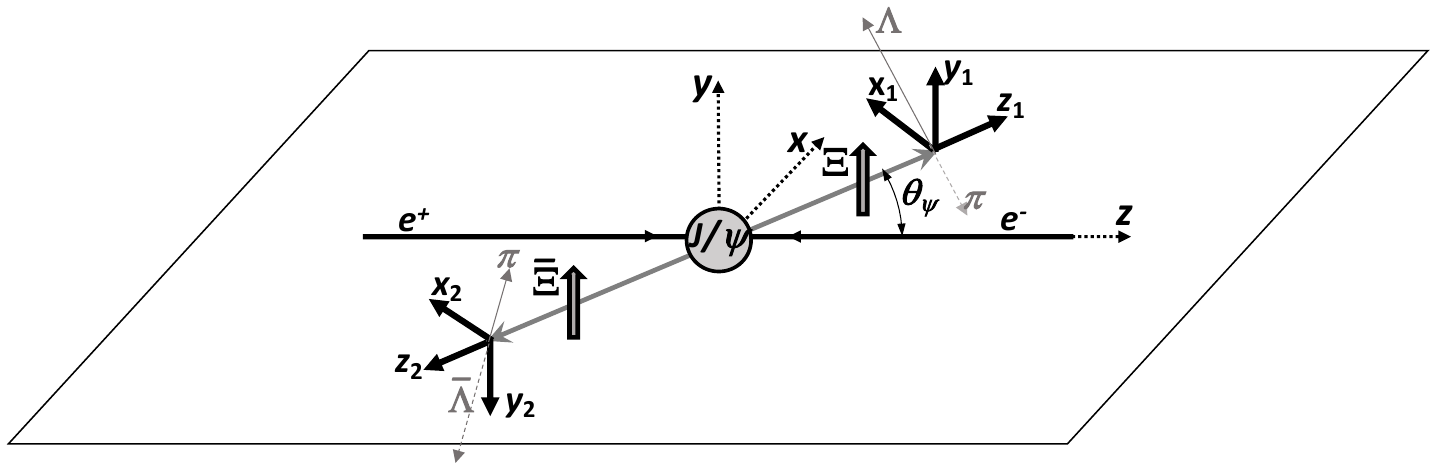}
    \caption{\footnotesize Coordinate systems for $\jpsi$$\rt$$\Xi\,\bar{\Xi}$ and $\Xi$$\rt$$\Lambda\pi$ and
      $\bar{\Xi}$$\rt$$\bar{\Lambda}\pi$ decays. The coordinate systems for $\Lambda$$\rt$$p\pim$ and
    $\bar{\Lambda}$$\rt$$\bar{p}\pip$ decays are not shown.}
    \label{fig:Jpsi-2-XiXibar}
  \end{center}
\end{figure}

The simplicity of the event topology belies a rather intricate analysis. Nine angular measurements are needed to completely
specify the kinematics of the event: the production angle, $\theta_{\psi}$, and a pair of angles, $\theta_Y~\&~\varphi_Y$ for
each of the four hyperon decay vertices (see Fig.~\ref{fig:Jpsi-2-XiXibar}).
The function that describes these highly correlated angular distributions is characterized by eight
parameters: $\alpha_{\psi}$~\&~$\Delta\Phi$ for the $\jpsi$$\rt$$\Xi\,\bar{\Xi}$ decay vertex, two parameters
for each $\Xi$ decay and one parameter for each $\Lambda$ decay (in which the proton polarization is not
measured). The measured quantities are expressed as a nine-component vector ${\boldsymbol \xi}$ and the
angular distribution by an eight-component vector ${\boldsymbol \omega}$:
\begin{eqnarray}
   \label{eqn:W}
  {\boldsymbol \xi} &=&
  (\theta_{\psi},\theta_{\Lambda},\varphi_{\Lambda},\theta_{\bar{\Lambda}},\varphi_{\bar{\Lambda}},\theta_{p},\varphi_{p},
    \theta_{\bar{p}},\varphi_{\bar{p}})\\
    {\boldsymbol \omega} &=&
  (\alpha_{\psi},\Delta\Phi,\alpha_{\Xi},\phi_{\Xi},\alpha_{\bar{\Xi}},\phi_{\bar{\Xi}},\alpha_{\Lambda},
    \alpha_{\bar{\Lambda}}).
\end{eqnarray}
Perotti~{\it et al.}~\cite{Perotti:2018wxm} provide a modularized expression for the nine dimensional probability
distribution that is in a convenient form for symbolic computing:

\begin{equation}
  \label{eqn:W}
  {\mathcal W}({\boldsymbol \xi}:{\boldsymbol \omega})=\sum^{3}_{\mu\nu=0}C_{\mu\nu}
  \bigg(\sum^3_{\mu'=0}a^{\Xi}_{\mu\mu'}a^{\Lambda}_{\mu'0}\bigg)
           \bigg(\sum^3_{\nu'=0}a^{\bar{\Xi}}_{\nu\nu'}a^{\bar{\Lambda}}_{\nu'0}\bigg),
\end{equation}
where $C_{\mu\nu}$ is  a 4$\times$4 matrix that contains the $\theta_{\psi}$-dependent $\Xi~\&~\bar{\Xi}$
polarizations and spin correlations, and each $a^Y_{\alpha\alpha^{\prime}}$ represents a 4$\times$4 matrix with
elements that contain expressions that involve  $\alpha_Y~\&~\phi_Y$ parameters and Wigner D-functions of the
$\theta_Y,\varphi_Y$ angular variables. The second subscript of the $a^{\Lambda/\bar{\Lambda}}_{\nu'0}$ terms are
zero because the $p$\,\&\,$\bar{p}$ polarizations are not measured. The sums in eqn.~\ref{eqn:W} involve a total of
256 expressions, of which 100 are nonzero.

\noindent
The  $C_{\mu\nu}$ spin-density matrix in eqn.~\ref{eqn:W} has the form
\begin{equation}
  \label{eqn:Cmunu}
  C_{\mu{\nu}}(\theta_{\psi},\alpha_{\psi},\Delta\Phi)=2(1+\alpha_{\psi}\cos^2\theta_{\psi})
  \begin{pmatrix}
          1         &   0     &   \mathcal{P}_y &   0   \\
          0         & C_{xx}  &       0        & C_{xz} \\
 -{\mathcal P}_{y}  &   0      &     C_{yy}     &   0   \\
 0         & -C_{xz}  &       0        & C_{zz} \end{pmatrix},
\end{equation}
where the 0$^{\rm th}$ row is the polarization four-vector\footnote{The polarization four vector is
   defined as $(1,\mathcal{P}_x,\mathcal{P}_y,\mathcal{P}_z)$.}
of the $\Xi$ in the $(x_1,y_1,z_1)$ coordinate system in Fig.~\ref{fig:Jpsi-2-XiXibar}; the  0$^{\rm th}$ column is
the same quantity for the $\bar{\Xi}$ in the $(x_2,y_2,z_2)$ system, and the 3$\times$3 submatrix $C_{ij}$ contains
the $\Xi$-$\bar{\Xi}$ spin-correlation coefficients. Parity conservation in the $\Xi\,\bar{\Xi}$ production process
requires that only $\mathcal{P}_y$ and $\bar{\mathcal{P}}_y$ are nonzero and $\mathcal{P}_y$=$\bar{\mathcal{P}}_y$
if referenced to the ($x,y,z$) $\jpsi$  rest frame coordinate system.\footnote{The minus sign
          in $C_{20}$=$-\mathcal{P}_y$ reflects the opposite
          directions of the $y_1$\,\&\,$y_2$ axes in Fig.~\ref{fig:Jpsi-2-XiXibar}.}

\begin{figure}[h!]
   \begin{center}
    \includegraphics[width=0.99\textwidth]{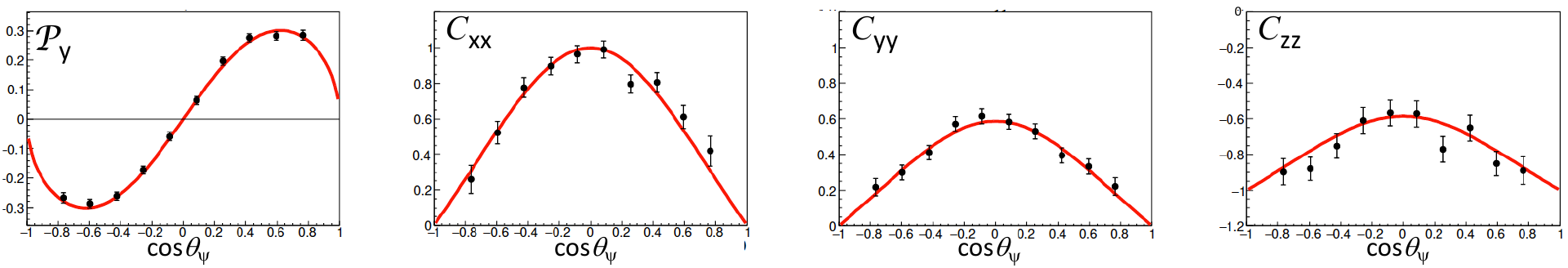}
    \caption{\footnotesize $\Xi$ polarization $\mathcal{P}_y$ and the $\Xi$-$\bar{\Xi}$ spin correlations  $C_{xx}$,
      $C_{yy}$ and $C_{zz}$ in quantum entangled $\jpsi\rt\Xi^-\bar{\Xi}^+$ events. (From BESIII~\cite{BESIII:2021ypr}.)}
    \label{fig:Xi-polarization}
  \end{center}
\end{figure}
\noindent
Figure~\ref{fig:Xi-polarization} show the $\cos\theta_\psi$-dependence of the $\Xi$ polarization and the $C_{xx}$,
$C_{yy}$ and $C_{zz}$ correlations coefficients for $\Xi^-\bar{\Xi}^+$ events. Note that although the average polarization is zero,
its rms value is nonzero at $\langle{\mathcal P}^\Xi_y \rangle_{\rm rms}$=\,22.3\%\,.  If the proton and antiproton directions
from secondary $\Lambda$ and $\bar{\Lambda}$ decays are not included in the fit, the 3$\times$3 $C_{ij}$ correlations do not apply,
$\phi_\Xi$ and $\phi_{\bar{\Xi}}$ could not be determined, and the $\alpha_\Xi$ and $\alpha_{\bar{\Xi}}$ measurement sensitivities
would be reduced by a factor equal to $\langle{\mathcal P}^\Xi_y\rangle_{\rm rms}$. With the inclusion
of the $p$\,\&\,$\bar{p}$ measurements, the $C_{ij}$ correlations come into play, $\phi_\Xi$~\&~$\phi_{\bar{\Xi}}$  are measured,
and the effective  $\Xi$ polarizations for the $\alpha_\Xi$~\&~$\alpha_{\bar{\Xi}}$ determinations become $\mathcal{P}^{\rm eff}_y$=1.
The corresponding overall sensitivity gain from the $\Lambda$ polarization measurements is equivalent to a twentyfold increase in
the number of events.

\begin{table}[h!]
  \footnotesize
  \caption{\label{tbl:BESIII-Xi-results}\footnotesize    Results from the BESIII
    $\Xi^0\bar{\Xi}^0$ analysis and the errors from their $\Xi^-\bar{\Xi}^+$ analysis.
  }
\begin{center}
\begin{tabular}{lccc}
\hline
Parameter  &$\Xi^0(\Lambda\piz)\bar{\Xi}^0(\bar{\Lambda}\piz)$~\cite{BESIII:2023drj}  &~&
                $\Xi^-(\Lambda\pim\bar{\Xi}^+(\bar{\Lambda}\pip)$~\cite{BESIII:2025abc}  \\
\hline
Number of events                &             327,305                     &~&        576,357                             \\
\hline
$\alpha_\psi$                   & $~~~0.514~     \pm 0.006  \pm 0.015$     &~&    $\ldots~~~\pm 0.0043  \pm 0.0035$    \\
$\Delta\Phi$\,(rad.)            & $~~~1.168~     \pm 0.019  \pm 0.018$     &~&   $\ldots~~~\pm 0.0161  \pm 0.0091$ \\
$\alpha_\Xi$                     & $-0.3750  \pm 0.0034 \pm 0.0016$     &~&    $\ldots~~~\pm 0.0026 \pm 0.0008$\\
$\alpha_{\bar{\Xi}}$              &  $~~~0.3790 \pm 0.0034   \pm 0.0021$   &~&   $\ldots~~~\pm 0.0026   \pm 0.0007$   \\
$\phi_\Xi$\,(rad.)               &  $~~~0.0051  \pm 0.0096   \pm 0.0018$  &~&    $\ldots~~~\pm 0.0072   \pm 0.0012$\\
$\phi_{\bar{\Xi}}$\,(rad.)        & $-0.0053   \pm 0.0097  \pm 0.0019$   &~&     $\ldots~~~\pm 0.0072  \pm 0.0016$ \\
$\alpha_\Lambda$                  &  $~~~0.7551  \pm 0.0052  \pm 0.0023 $   &~&    $\ldots~~~\pm 0.0038  \pm 0.0014 $\\
$\alpha_{\bar{\Lambda}}$           & $-0.7448  \pm 0.0052   \pm 0.0017 $   &~&   $\ldots~~~\pm 0.0038   \pm 0.0016 $ \\
\hline
$\xi_P$$-$$\xi_S$\,(rad.)        &  $(~~~0.0 \pm 1.7 \pm 0.2)\times 10^{-2}$  &~&   $(\ldots~\pm 1.2 \pm 0.2)\times 10^{-2}$ \\
$\delta_P$$-$$\delta_S$\,(rad.)  & $(-1.3 \pm 1.7 \pm 0.4)\times 10^{-2}$   &~&  $(\ldots~\pm 1.2 \pm 0.2)\times 10^{-2}$\\
\hline
$A^{\Xi}_{\CPa}$                  & $(-5.4  \pm 6.5 \pm 3.1)\times 10^{-3}$  &~&   $(\ldots~\pm 4.8 \pm 1.4)\times 10^{-3}$   \\
$\Delta\phi^{\Xi}_{\CPa}$\,(rad.) & $(-0.1  \pm 6.9 \pm 0.9)\times 10^{-3}$   &~&   $(\ldots~\pm 5.1 \pm 1.0)\times 10^{-3}$  \\
$A^{\Lambda}_{\CPa}$               & $(~~~6.9  \pm 5.8 \pm 1.8)\times 10^{-3}$   &~&    $(\ldots~\pm 4.3 \pm 1.4)\times 10^{-3}$   \\
\hline
$\langle\alpha_\Xi\rangle$       & $-0.3770 \pm 0.0024 \pm 0.0014$          &~&    $\ldots~~~\pm 0.0018 \pm 0.0005$  \\
$\langle\phi_\Xi\rangle$\,(rad.) & $~~~0.0052 \pm 0.0069 \pm 0.0016$       &~&    $\ldots~~~\pm 0.0050 \pm 0.0009$  \\
$\langle\alpha_\Lambda\rangle$    & $-0.7499 \pm 0.0029 \pm 0.0013$        &~&   $\ldots~~~\pm 0.0021 \pm 0.0011$    \\
\hline
\hline
 \end{tabular}
\end{center}
\end{table}

The results of BESIII analyses of the 321\,k $\Xi^0\bar{\Xi}^0$ and the errors of 576\,k $\Xi^-\bar{\Xi}^+$ event sample are summarized
in Table~\ref{tbl:BESIII-Xi-results}.  (The $\Xi^-\bar{\Xi}^+$ analysis is complete and the results are known within the BESIII
collaboration but, until they are published, only the errors have been released to the public.) Of particular note are the BESIII
measurements of $\phi_{\Xi}$ and $\phi_{\bar{\Xi}}$ that, in degrees, are
\begin{center}
  \footnotesize
  \begin{tabular}{lccc}
  Channel           &      $\phi_\Xi$     &    $\phi_{\bar{\Xi}}$     &     $\xi_P$$-$$\xi_S$    \\
  \hline
  $\Xi^0\bar{\Xi}^0$ & $~~~0.29^\circ\pm 0.55^\circ \pm 0.10^\circ$ &   $-0.30^\circ\pm 0.56^\circ \pm 0.11^\circ$ &
       $~~~0.00^\circ \pm 0.97^\circ  \pm 0.11^\circ$             \\  
$\Xi^-\bar{\Xi}^+$ & $~~~~~\ldots^\circ\pm 0.41^\circ \pm 0.07^\circ$ &   $~~~~~\dots^\circ\pm 0.41^\circ \pm 0.09^\circ$ &
       $~~~~~~\ldots^\circ \pm 0.68^\circ  \pm 0.11^\circ$            
\end{tabular}
\end{center}
and, when the $\Xi^0$ and $\Xi^-$ $\xi_P$$-$$\xi_S$ values are combined,\footnote{For eqns.~\ref{eqn:xi_p-xi_S-limit}
        and \ref{eqn:delta_p-delta_S-limit}, the $\Xi^-\bar{\Xi}^+$ and $\Xi^0\bar{\Xi}^0$
        analyses central values were assumed to be that same.}
we get a sub-1$^\circ$ upper limit:
\begin{equation}
  \label{eqn:xi_p-xi_S-limit}
|\xi_P-\xi_S| = 0.00^\circ \pm 0.56^\circ \pm 0.11^\circ~~~<0.94^\circ~~~{\rm (90\%~C.L.)}\,,
\end{equation}
that is limited by statistical errors.  A similar average on the $\delta_P$$-$$\delta_S$ strong phase difference
results in very nearly the same upper  limit
\begin{equation}
  \label{eqn:delta_p-delta_S-limit}
|\delta_P-\delta_S| <1^\circ~~~{\rm (90\%~C.L.)}\,,
\end{equation}
which is much smaller than the $\sim$8.8$^\circ$ estimate given in ref.~\cite{Huang:2017bmx} that was mentioned above.
According to the definition given above in eqn.~\ref{eqn:ACP-asymmetry}, this low value of $\delta_P$$-$$\delta_S$  means that
$|A^\Xi_{\CPa}|$ will be $\lesssim$0.02($\xi_P$$-$$\xi_S$) and in all likelihood too small to be a useful probe for \C\Par~violations.

On the other hand, according to eqn.~\ref{eqn:Lambda-pol} and Fig.~\ref{fig:Xi_and_Lambda-decays}a, the $\Lambda$ hyperons produced
in $\Xi$$\rt$$\Lambda\pi$ decays have a polarization $|\mathcal{P}_\Lambda|$\,$\approx$\,$|\alpha_\Xi|$\,=\,0.38, that is twice the
$\langle{\mathcal P}^{\Lambda}_y\rangle_{\rm rms}$=18.7\% polarization of lambdas that are produced directly via the
$\ee$$\rt$$\jpsi$$\rt$$\Lambda\bar{\Lambda}$ reaction. 
As a result, it is interesting to compare  measurements of $A^\Lambda_{\CPa}$ listed in Table~\ref{tbl:BESIII-Xi-results} from the
327\,k event $\Xi^0\bar{\Xi}^0$ and 576\,k event $\Xi^-\bar{\Xi}^+$ analyses with a  BESIII result based on an analysis of
3.23\,M $\jpsi$$\rt$$\Lambda\bar{\Lambda}$ events~\cite{BESIII:2022qax}:
\begin{eqnarray}
          \Xi^0\bar{\Xi}^0:~~~ A^\Lambda_{\CPa}&=& (~~~6.9 \pm5.8 \pm 1.8)\times 10^{-3}~~~~327\,{\rm k}~{\rm evts}\\
          \Xi^-\bar{\Xi}^+:~~~ A^\Lambda_{\CPa}&=& (~\ldots~\pm4.3 \pm 1.4)\times 10^{-3}~~~~576\,{\rm k}~{\rm evts}\\
  \Lambda^0\bar{\Lambda}^0:~~~ A^\Lambda_{\CPa}&=& (-2.5 \pm4.6 \pm 1.2)\times 10^{-3}~~~~3.23\,{\rm M}~{\rm evts}.
\end{eqnarray}
Despite having much smaller data samples, the $\jpsi$$\rt$$\Xi\,\bar{\Xi}$ event samples provide independent measurements of
$A^\Lambda_{\CPa}$ that are similar in precision to those based on $\jpsi$$\rt$$\Lambda\bar{\Lambda}$ events and with a somewhat
different set of systematic errors.

\subsection{Future prospects}

The eqn.~\ref{eqn:xi_p-xi_S-limit} limit, $|\xi_P-\xi_S|$$<$0.94$^\circ$  leaves considerable room for the influence of possible
new physics in $\Xi$  hyperon decay processes to appear before  the expected level  SM \C\Par~violations is reached. The most recent
calculation of the SM values for the $\xi_P$ and $\xi_S$ $CP$-violating phases found their difference to be~\cite{Tandean:2002vy}
\begin{equation}
  \label{eqn:SMxiP-xiS}
  (\xi_P -\xi_S)_{\rm SM}= -(1.4\pm 1.2)A^2\lambda^5\bar{\eta}=-0.01^{\circ}\pm 0.01^{\circ},
  \end{equation}
    where $A^2\lambda^5\bar{\eta}$ is the imaginary part of the $V^*_{td}V^{~}_{ts}$ CKM matrix element product in
    the  Wolfenstein parameterization~\cite{Wolfenstein:1983yz}.  This is a factor of a hundred below the statistical precision of the
    above-described BESIII  measurement.

But not the systematic error, which is only ten times the SM value for $\xi_P$$-$$\xi_S$, despite the fact that  \C\Par~studies were
not considered during the design of the general-purpose BESIII detector. This reflects the above-mentioned features of the
$\jpsi$$\rt$$\Xi\,\bar{\Xi}$ events that make them a robust platform for \C\Par~violation studies.  A detector at a future facility could be
more closely tailored to the requirements of \C\Par~studies, and, since BESIII's quoted 0.1$^\circ$ systematic errors are, for the most part,
statistical errors associated with the   numbers of control sample events that were used for their evaluation, these would be smaller for
larger $\jpsi$ event samples. Thus, it  seems reasonable to expect that a specialized facility capable of producing and handling multi-trillion
$\jpsi$ event samples could have an  $\xi_P$$-$$\xi_S$ measurement sensitivity that is near the 0.01$^\circ$ SM level.

\section{Tests of the \C\Par\T~invariance with $\boldsymbol{\jpsi}$ decays}
The \C\Par\T~~theorem ~\cite{Schwinger:1951xk}
states that any quantum field theory  that is {\it Lorentz invariant}, has {\it local point-like interaction vertices},
and is {\it hermitian} ({\it i.e.}, conserves probability) is invariant under the combined operations of
\C, \Par~and \T. Since the three quantum field theories that make up the Standard Model---QED, QCD, and
Electroweak theory---all satisfy these criteria, \C\Par\T~symmetry is a fundamental and inescapable prediction of the theory.
Any deviation from \C\Par\T~invariance would be unambiguous evidence for new, beyond the
Standard Model physics.  Given the central role that the \C\Par\T~theorem plays in our understanding of the most basic
properties of space and time, it is essential that its validity should be tested at the highest level of sensitivity
that experimental technology permits.

Fortunately, the $\Kz$-$\Kzbar$ mixing process coupled with the requirements of unitarity provide sensitive ways to search
for violations of the \C\Par\T~theorem.  This is embodied in a remarkable equation called the
{\it Bell-Steinberger relation} that unambiguously relates a fundamental \C\Par\T~violating amplitude called $\delta$ to
well defined measurable quantities.  The relation is exact and, unlike most other experimental tests of the Standard
Model, there are no loose ends such as long-distance QCD-corrections or theoretical assumptions about anything other
than unitarity and time independence of the Hamiltonian. If a nonzero value of $\delta$ is established, the only possible
interpretations are that either the \C\Par\T~theorem is invalid,  or unitarity is violated, or that kaon mass or lifetime
changes with time.

The measurement involves studies of the proper-time-dependence of the  decay rates of large numbers of neutral kaons that
have well established  ({\it i.e., tagged}) strangeness quantum numbers at the time of their production $\Kz(\tau)$ for
\Str=+1 and $\Kzbar(\tau)$ for \Str=$-$1.  The most sensitive limits to date are from experiments in the 1990s that used data
samples that contained tens of millions of $\Kz$ and $\Kzbar$ decays. A $\sim$10$^{12}$-event sample of $\jpsi$  decays
will contain $\sim$2\,B $\jpsi$$\rt$$\Km\pip\Kz$ events  and an equivalent number of $\jpsi$$\rt$$\Kp\pim\Kzbar$ events where
the neutral kaon decays to $\pipi$, and $\sim$1\,B events for each mode in which the kaon decays to $\piz\piz$. In
these reactions, the initial strangeness of the neutral kaon is tagged by the sign of the accompanying charged-kaon's
electric charge: a $\Km$ tags an \Str=+1 $\Kz(\tau)$ and  a $\Kp$ tags an \Str=$-$1 $\Kzbar(\tau)$. These data samples
would support a repeat of previous measurements with $\sim$fifty-times larger data samples. 

In this talk I discuss:

\vspace{1mm}
\noindent
{\it i})~why we expect \C\Par\T~to be violated somewhere below the Planck mass-scale;

\vspace{1.5mm}
\noindent
{\it ii})~why the neutral kaon system is well suited for tests of the \C\Par\T~theorem;

\vspace{1.5mm}
\noindent
{\it iii})~the quantum-mechanics of neutral kaons with restrictions on \C\Par\T~ relaxed;

\vspace{1.5mm}
\noindent
{\it iv})~estimated experimental sensitivities with 10$^{12}$ $\jpsi$ decays;

\vspace{1.5mm}
\noindent
{\it v})~applications of the Bell-Steinberger relation at a specialized $\jpsi$ facility.

\subsection{\C\Par\T~and the Theory of Everything}

One of the requirements for a \C\Par\T-invariant theory is that it is {\it local}, which means that the
couplings at each vertex occurs at a single point in space-time. But theoretical physics has always had
troubles with point-like quantities. For example, the classical Coulombic self-energy of the electron is
\begin{equation}
  W_e=\frac{e^2}{4\pi\eps_0 r_e},
\end{equation}
which diverges for $r_e$$\rt$0.  The {\it classical radius of the electron}, {\it i.e.}, the value of $r_e$
that makes $W_e$=\,$m_ec^2$, is $r^{\rm c.r.e.}_e$=\,2.8$\times$10$^{-13}$~cm (2.8 fermis),  which is three times the
radius of the proton, and $\sim$3000 times larger than the experimental upper limit on the electron radius, which
is of $\mathcal{O}(10^{-16}\,{\rm cm})$~\cite{ZEUS:2003eqd}. Infinities associated with point-like objects persist
in quantum field theories, where they are especially troublesome. In second- and higher-order perturbation theory, all
of the diagrams that have virtual-particle loops involve integrals over all possible configurations of the 
loops that conserve energy and momentum.  Whenever two of the point-like vertices coincide,
the integrands become infinite and cause the integrals to diverge.

\begin{figure}[h!]
\centering
\includegraphics[width=0.6\textwidth]{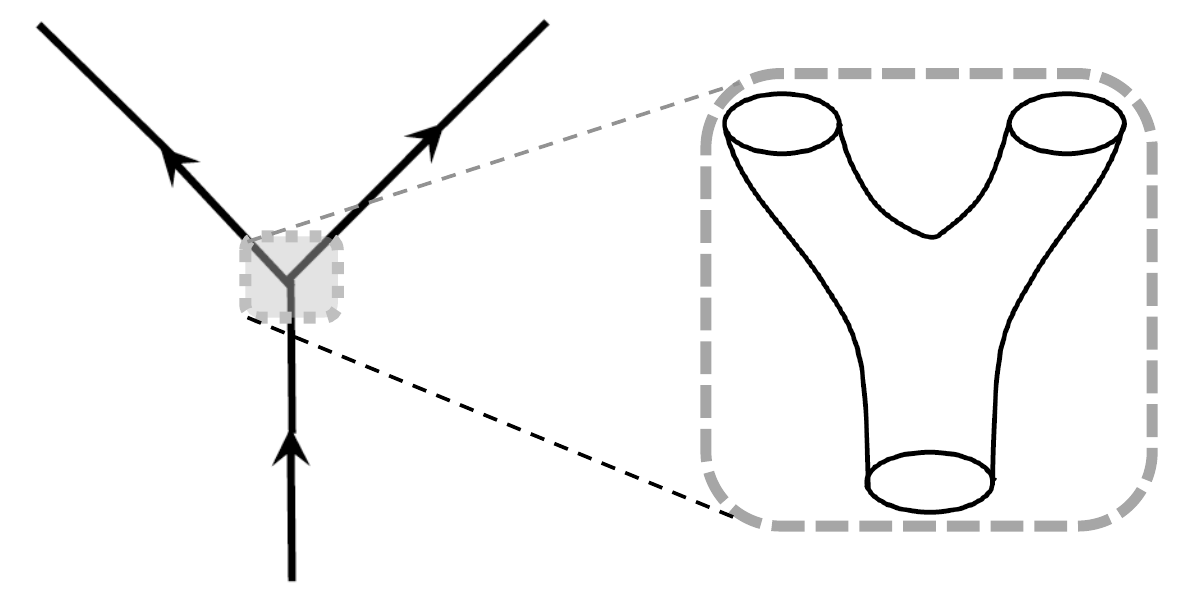}
\caption{\footnotesize In string theories, elementary particles are tiny loops
  of oscillating strings with no point-like vertices and their associated infinities.}
 \label{fig:String-vertex}
 \end{figure}

In the QED, QCD and Electroweak quantum field theories that make up the Standard Model, these infinities are
removed by the well established methods of renormalization~\cite{Bethe:1947id,Dyson:1949bp,Wilson:1973jj},
a technique that involves the subtraction of two divergent quantities---where one of the divergences is associated with
an established observable (like the free electron charge, or mass)---to produce a finite result,  In all
three of these theories, the perturbation expansions are in increasing powers of a dimensionless coupling
strength,  $\alpha_{\rm EW}$\,=\,$\sqrt{2}M^2_WG_F/\pi$,\footnote{Specifically
       not just $G_F$, which has dimensions of mass$^{-2}$.}
        $\alpha_{\rm QED}$, and $\alpha_s$.
As a result of this, relations that exist between different orders of the
perturbation expansion reduce the number of observed quantities that are needed to subtract off divergences.
In QED, for example, there are only two, the electron's mass, and charge.   However, in quantum theories of gravity, where a
massless spin=2 {\it graviton} plays the role of the photon in QED, the  expansion constant is Newton's gravitational constant
with units m$^3$/kg$\cdot$s$^2$. Because of this, every order in the perturbation expansion has different dimensions and, thus,
needs different observables to carry out the subtractions at each step. Thus a complete renormalization would
require an infinite number of observed quantities to complete the renormalization, with the end effect that a quantum theory
of gravity with point-like vertices would, in principle, be {\it non-renormalizable}\cite{Weinberg:1980kq}. As a result,
a viable theory of everything will have to include a mechanism for avoiding these divergences by smearing out the
vertices, for example by treating particles as loops of string as in Fig.~\ref{fig:String-vertex}, and this
{\it non-locality} would eliminate one of the necessary conditions for  \C\Par\T~invariance. In  such a scenario, the
Theory of Everything, which, by definition, would have to include gravity,\footnote{Even cavemen were aware of gravity.}
would not be local and include \C\Par\T~violations at some mass scale.

Although non-renormalizable higher-order perturbative effects do not show up at mass scales below the
Planck mass ($M_P$\,=\,$\sqrt{\hbar c/G_N}$\,=1.22$\times$10$^{19}$GeV), the fact that this problem exists dispels the notion
that there is something especially sacred about \C\Par\T-invariance; there is no fundamental principle that prevents it from
  being violated at a lower mass scale.  Because of its close connection with the fundamental assumptions of the Standard Model,
  stringent experimental tests of \C\Par\T~invariance should have high priority

\subsection{Neutral $\boldsymbol{K}$ mesons and tests of the \C\Par\T~~theorem}

The main consequences of \C\Par\T~symmetry are that particle and antiparticle masses and lifetimes are equal.
Since lifetime differences can only come from on-mass-shell intermediate states and do not probe short-distance
high-mass physics, these are unlikely to exhibit any \C\Par\T-violating asymmetry. Instead, our focus here is on the
possibility that particle and antiparticle masses may be different.

\begin{figure}[!tbp]
\centering
\includegraphics[width=0.99\textwidth]{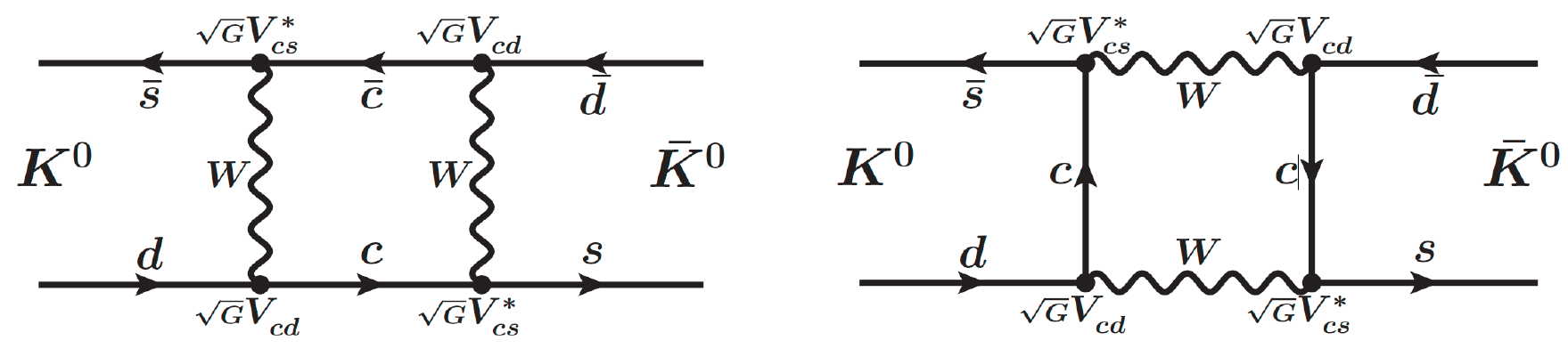}
\caption{\footnotesize The box diagrams for the short-distance contributions to  $\Kz$-$\Kzbar$
mixing.}
 \label{fig:k-mix_c-quark-KM}
  \end{figure}

The particles with the best measured masses are the stable electron and proton, and, according the
PDG~2020 tables~\cite{Zyla:2020zbs}:
\begin{eqnarray}
    \label{eqn:eebar-limit}
  |m_{e^+}-m_{e^-}|&<&4\times 10^{-9}~{\rm MeV}\\
    \label{eqn:ppbar-limit}
  |m_{\bar{p}}-m_{p}|&<&7\times 10^{-7}~{\rm MeV}.
\end{eqnarray}
However, these limits do not provide the best tests of \C\Par\T; the most stringent experimental restriction on
\C\Par\T~violation comes from the difference between the $\Kzbar$ and $\Kz$ masses:
\begin{equation}
  \label{eqn:KKbar-limit}
  |M_{\Kzbar}-M_{\Kz}|<5\times 10^{-16}~{\rm MeV},
\end{equation}
which is $7$-$9$~orders of magnitude more strict than those from the electron and proton mass measurements,
even though the value of $M_{\Kz}$ itself is only known to~$\pm 13$~keV. This is because the
Fig.~\ref{fig:k-mix_c-quark-KM} diagrams, taken together with the quantum mechanics of $\Kz$-$\Kzbar$ mixing, map the
$M_{\Kzbar}$$-$$M_{\Kz}$ difference into the quantities $\Delta M$\,=\,$M_{\KL}$$-$$M_{\KS}$, which is
3.5$\times$10$^{-12}$~MeV and 14 orders of magnitude lower than $M_{\Kzbar}$ or $M_{\Kz}$, and  
$\Delta\Gamma$=\,$\Gamma_{\KS}$$-$$\Gamma_{\KL}$$\approx$\,7.4$\times$10$^{-12}$~MeV, which is, coincidently,
$\approx$2$\times\Delta M$.

\subsection{The neutral kaon mass eigenstates with no \C\Par\T-invariance related restrictions} 

Any arbitrary neutral kaon state in its rest frame can be expressed as a linear combination of the strangeness
eigenstates, $\ket{\psi}$=\,$\alpha_1(\tau)\ket{\Kz}$\,+\,$\alpha_2(\tau)\ket{\Kzbar}$, where $\alpha_1(\tau)$ and
$\alpha_2(\tau)$ are complex functions that are normalized as $|\alpha_1(0)|^2$+\,$|\alpha_2(0)|^2$=\,1. In the
Wigner-Weisskopf formulation of quantum mechanics for exponentially decaying systems, the time dependence is included
by using $\alpha_i(\tau)$\,=\,$\alpha^k_ie^{-i\lambda_k\tau}$ where $\alpha^k_i$ are complex constants and the
Schr\"{o}dinger equation in the form
\begin{equation}
  \label{eqn:time-dep-schrod}
  \boldsymbol{\Ham}\psi(\tau)=\big(\boldsymbol{\Mass}-\frac{i}{2}\Gmm\big)\boldsymbol{\Psi}_k e^{-i\lambda_k\tau}
  =i\frac{d\psi_K}{d\tau}=\lambda_k\boldsymbol{\Psi}_ke^{-i\lambda_K\tau},
\end{equation}
where the Hamiltonian $\boldsymbol{\Ham}$, the {\it mass matrix} $\boldsymbol{\Mass}$ and the {\it decay matrix}
$\Gmm$ are 2$\times$2 matrices and $\boldsymbol{\Psi}_k$ is a two-dimensional spinor in ($\Kz,\Kzbar$) space:
\begin{equation}
\boldsymbol{\Mass}=\begin{pmatrix} M_{11} & M_{12}\\ M^*_{12} & M_{22}\end{pmatrix},~~~~~
\Gmm=\begin{pmatrix} \Gamma_{11} & \Gamma_{12}\\ \Gamma^*_{12} & \Gamma_{22}\end{pmatrix}~~~~{\rm and}~~~~
\boldsymbol{\Psi}_k=\begin{pmatrix} \alpha^k_1 \\ \alpha^k_2\end{pmatrix}.
\end{equation}
Here $\boldsymbol{\Mass}$ and $\Gmm$ are hermitian, which is ensured by setting $M_{21}$\,=\,$M^*_{12}$ and
$\Gamma_{21}$\,=\,$\Gamma^*_{12}$, while $\boldsymbol{\Ham}$, which describes decaying particles and doesn't conserve
probability, is not hermitian.  Symmetry under \C\Par~requires $M_{12}$ to be real; \C\Par\T~invariance requires
$M_{11}$=\,$M_{22}$ and $\Gamma_{11}$=\,$\Gamma_{22}$. The mass eigenvalues and eigenstates for the most general case,
{\it i.e.}, with neither \C\Par~nor \C\Par\T~symmetry are~\cite{Bigi:2000yz,Schubert:2014ska}
\begin{eqnarray}
  \label{eqn:mass-eigenstates-noCPT}
 \lambda_S=M_S-\rootionehalf\Gamma_S;~~~\ket{\KS}={\textstyle \frac{1}{\sqrt{2(1+|\eps_S|^2)}}}
  \big[\big(1+{\eps_S}\big)\ket{\Kz}
      +\big(1-{\eps_S}\big)\ket{\Kzbar}\big]\,&&\\
  \nonumber
 \lambda_L=M_L-\rootionehalf\Gamma_L;~~~\ket{\KL}={\textstyle \frac{1}{\sqrt{2(1+|\eps_L|^2)}}}
    \big[\big(1+{\eps_L}\big)\ket{\Kz}
               -\big(1-{\eps_L}\big)\ket{\Kzbar}\big],&&
\end{eqnarray}  
\noindent
where $\eps_S$\,=\,$\eps$+$\delta$, $\eps_L$=\,$\eps$$-$$\delta$ and $\eps$\,\&\,$\delta$ can be expressed in terms of
$\Delta M$\,$\equiv$\,$M_L$$-$$M_S$ and $\Delta\Gamma$$\equiv$\,$\Gamma_S$$-$$\Gamma_L$ as
\begin{equation}
  \label{eqn:def-eps}
  \eps=\frac{-i\Im M_{12}-{\textstyle \frac{1}{2}}\Im \Gamma_{12}}{{\Delta M + \textstyle \frac{i}{2}}\Delta\Gamma},
\end{equation}
which is the familiar neutral kaon mass-matrix \C\Par~violation parameter, and
\begin{equation}
  \label{eqn:deltaCPT}
  \delta=\frac{(M_{\Kzbar}-M_{\Kz})-i(\Gamma_{\Kzbar}-\Gamma_{\Kz})/2}{2\Delta M -i\Delta\Gamma},
\end{equation}
is the equivalent dimensionless parameter that characterizes mass-matrix \C\Par\T~violation parameter.  A unique and important
feature of this scenario is that the $\KL$ and $\KS$
eigenstates are not orthogonal; according to eqn.\,\ref{eqn:mass-eigenstates-noCPT},
\begin{equation}
  \label{eqn:KLKS-CPT}
  \braket{\KS}{\KL}=2\Re \eps -2i\Im \delta,
\end{equation}
where here, and in (most of) the following, second-order terms in $|\eps|$ are dropped.

\subsubsection{Properties of $\boldsymbol{\eps}$ and $\boldsymbol{\delta}$}

In a commonly used phase convention, $\Im\Gamma_{12}$ is defined to be zero, in which case the
phase of $\eps$ is directly related to the well measured quantities\footnote{We use\,\cite{Workman:2022ynf}
  $\Delta M$\,=\,$(0.5289\pm 0.0010)$$\times$10$^{10}$s$^{-1}$\,\&\,$\Delta\Gamma$=\,$(1.1149\pm 0.0005)$$\times$10$^{-10}$s$^{-1}$.}
$\Delta M$ and $\Delta\Gamma$ via
\begin{equation}
  \phi_{\rm SW} = \tan^{-1}\bigg(\frac{2\Delta M}{\Delta\Gamma}\bigg)=43.5^\circ\pm 0.1^\circ,
  ~~~\leftarrow~{\rm the~{\it Superweak~phase}}
\end{equation}
that, since $\Delta M$\,$\approx$\,$\Delta\Gamma/2$, is very nearly 45$^\circ$.
Using this approximate relation for $\phi_{\rm SW}$, we can rewrite  the eqn.\,\ref{eqn:deltaCPT} expression for $\delta$ as
\begin{equation}
  \label{eqn:deltaCPT-modified}
  \delta \approx \frac{i(M_{\Kzbar}-M_{\Kz})+(\Gamma_{\Kzbar}-\Gamma_{\Kz})/2}{2\sqrt{2}\Delta M}e^{i\phi_{\rm SW}}
     = (i\delta_{\perp} + \delta_{\parallel})e^{i\phi_{\rm SW}}
  \end{equation}
where the real quantities $\delta_{\perp}$
and $\delta_{\parallel}$ are defined as
\begin{equation}
\label{eqn:delta-perp-par-def}
  \delta_{\perp} \approx \frac{M_{\Kzbar}-M_{\Kz}}{2\sqrt{2}\Delta M}~~~~~{\rm and}~~~~
   \delta_{\parallel} = \frac{(\Gamma_{\Kzbar}-\Gamma_{\Kz})}{4\sqrt{2}\Delta M}.
\end{equation}
In the complex plane, $\delta_{\perp}$, which is the short-distance component of $\delta$, is perpendicular
to $\eps$, and $\delta_{\parallel}$, the long-distance component, is  parallel to $\eps$.  Thus, a signature for a nonzero
short-distance \C\Par\T~violation would be a difference between the phase of $\eps_L$ and $\phi_{\rm SW}$ by an amount
\begin{equation}
\Delta\phi^{\mathcal{CPT}}=\phi_{\eps_L}-\phi_{\rm SW}=\frac{\delta_{\perp}}{|\eps|},
\end{equation}
which means that
\begin{eqnarray}
  \label{eqn:DeltaM_veltaPhi}
  &~&M_{\Kzbar}-M_{\Kz}=2\sqrt{2}|\eps|\Delta M\Delta\phi^{\mathcal{CPT}}\\
  \nonumber
  &~&~~~~~~~~~~~~~~~~~~~~~~=3.8\times 10^{-16}{\rm (MeV/deg)}\times\Delta\phi^{\mathcal{CPT}}.
\end{eqnarray}
The tiny proportionality constant between the \C\Par\T-violating $\Kz$-$\Kzbar$ mass difference
and  the $\Delta\phi^{\mathcal{CPT}}$  observable quantity is the reason the
eqn.\,\ref{eqn:KKbar-limit} limit on $M_{\Kzbar}$$-$$M_{\Kz}$ is many orders of magnitude more stringent than those for
$m_{\bar{p}}$$-$$m_{p}$ (eqn.\,\ref{eqn:ppbar-limit})    amd $m_{\bar{e}}$$-$$m_{e}$ (eqn.\,\ref{eqn:eebar-limit}).
This is the ``magic'' of the neutral kaon system and doesn't apply to any other particles.

\subsection{Interference measurements of the $\boldsymbol{\phi_{+-}}$ and $\boldsymbol{\phi_{00}}$ phases}

A simulated $\jpsi$$\rt$$\Km\pip\Kz(\tau)$; $\Kz(\tau)$$\rt$$\pipi$ event in the BESIII detector is shown in
Fig.~\ref{fig:k0kpi-Wu-Yang}a. Although it looks superficially like a $\Km\pip\KS$ event, the neutral kaon is not
a $\KS$ mass eigenstate and does not have a simple exponential decay curve. Instead, the neutral kaons produced in these
reactions are coherent mixtures of interfering $\ket{\KS}$ and $\ket{\KL}$ eigenstates:
\begin{figure}[b!]
\centering
\includegraphics[width=0.8\textwidth]{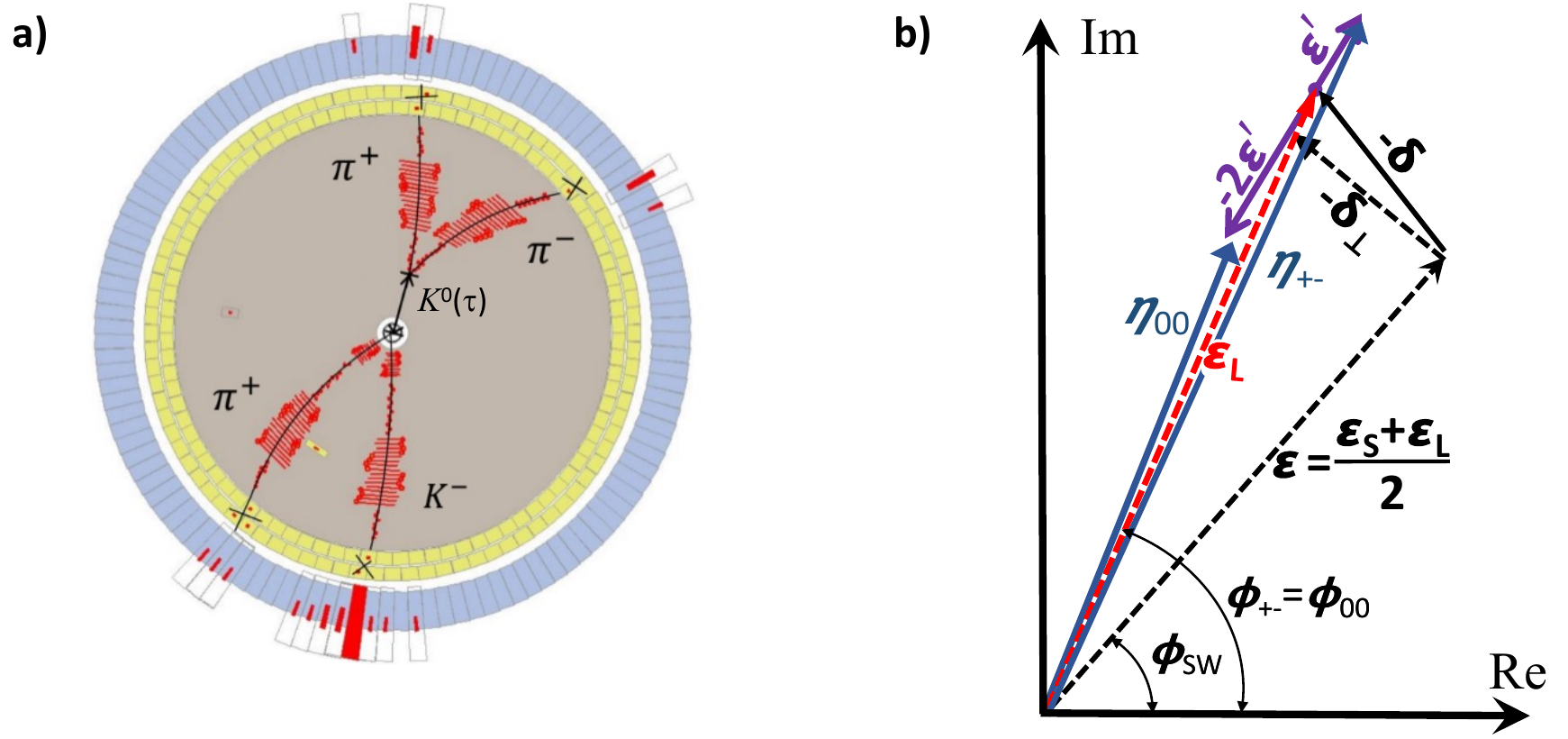}
\caption{\footnotesize {\bf a)}  A simulated $\jpsi$$\rt$$K^-\pip\Kz(\tau)$; $\Kz(\tau)$$\rt\pipi$ event in
  the BESIII detector.
  {\bf b)}  The relative arrangements on the complex plane of of the complex quantities discussed in the text.
  Here, for display purposes, the magnitudes of $\delta$ and $\epsp$ relative to $\eps$ are exaggerated.}
 \label{fig:k0kpi-Wu-Yang}
 \end{figure}
\begin{eqnarray}
  \ket{\Kz (\tau)}&=&{\textstyle \frac{1}{\sqrt{2}}}\big[(1+\eps_L)\ket{\KS}e^{-i\lambda_s\tau}
                 +(1+\eps_S)\ket{\KL}e^{-i\lambda_L\tau}\big]\\
  \nonumber
  \ket{\Kzbar (\tau)}&=&{\textstyle \frac{1}{\sqrt{2}}}\big[(1-\eps_L)\ket{\KS}e^{-i\lambda_s\tau}
                 -(1-\eps_S)\ket{\KL}e^{-i\lambda_L\tau}\big],
\end{eqnarray}
and have decay curves with opposite-sign $\KS$-$\KL$ interference terms:
\begin{equation}
  \begin{bmatrix} R(\tau)_{\Kz\rt\pi\pi}\\ \bar{R}(\tau)_{\Kzbar\rt\pi\pi}\end{bmatrix}
  \propto(1\mp 2\Re\eps_L)\big[e^{-\Gamma_S\tau}+|\eta_{j}|^2e^{-\Gamma_L\tau}
        \nonumber
        \pm 2|\eta_{j}|e^{-\bar{\Gamma}\tau}\cos\big(\Delta M\tau - \phi_{j}\big)\big],
\end{equation}
where $\bar{\Gamma}$=\,$(\Gamma_S$+$\Gamma_L)/2$\,$\approx$\,$\Gamma_S/2$  and
and $|\eta_{j}|$ and $\phi_j$
are the magnitudes and phases of $\eta_{+-}$ and $\eta_{00}$:
\begin{equation}
  \eta_{+-} = \frac{\braket{\pipi}{\KL}}{\braket{\pipi}{\KS}}=\eps-\delta+\epsp~~~~{\rm and}~~~~
  \eta_{00} = \frac{\braket{\piz\piz}{\KL}}{\braket{\piz\piz}{\KS}}=\eps-\delta-2\epsp.
  \end{equation}
Here $\epsp$=$\braket{\pi\pi}{\Ktwo}/\braket{\pi\pi}{\Kone}$ is the ratio of the direct-\C\Par-violating
amplitude of the \C\Par-odd $\ket{\Ktwo}$\,=$\smallrootonehalf\big(\ket{\Kz}$$-$$\ket{\Kzbar}\big)$ eigenstate
to \C\Par-even $\pi\pi$ final states to that for the \C\Par-allowed $\pi\pi$ decays of its \C\Par-even counterpart:
$\ket{\Kone}\,$=$\smallrootonehalf\big(\ket{\Kz}$+$\ket{\Kzbar}\big)$. Measurements~\cite{NA48:2002tmj,Abouzaid:2010ny}
have established that $\epsp$ is much smaller than $\eps$:
\begin{equation}
  \Re(\epsp/\eps) =(1.66\pm 0.23)\times 10^{-3},
\end{equation}
and its phase is determined from the analysis of low-energy $\pi\pi$ scattering measurements to be
$\phi_{\epsp}$=\,42.3$^\circ\pm$1.7$^\circ$~\cite{Colangelo:2001df} and equal, within errors, to $\phi_{\rm SW}$. This rather
remarkable coincidence\footnote{$\phi_{\epsp}$ is a long-distance QCD effect while
  $\phi_{\rm SW}$ is due to short-distance EW physics.}
means that $\epsp$ and $\eps$ are very nearly parallel, and the combined effect of this with its small
magnitude  reduces the impact of the uncertainty of the magnitude of $\epsp$ on $\delta\phi^{\mathcal{CPT}}$ measurements to negligible,
$\mathcal{O}(0.01^\circ$), levels. The arrangement of the various quantities discussed above on the complex plane is indicated in
Fig.~\ref{fig:k0kpi-Wu-Yang}b.

\subsubsection{Estimated measurement sensitivity with $\boldsymbol{10^{12}~\jpsi}$-decays}

\begin{figure}[h!]
\centering
\includegraphics[width=0.95\textwidth]{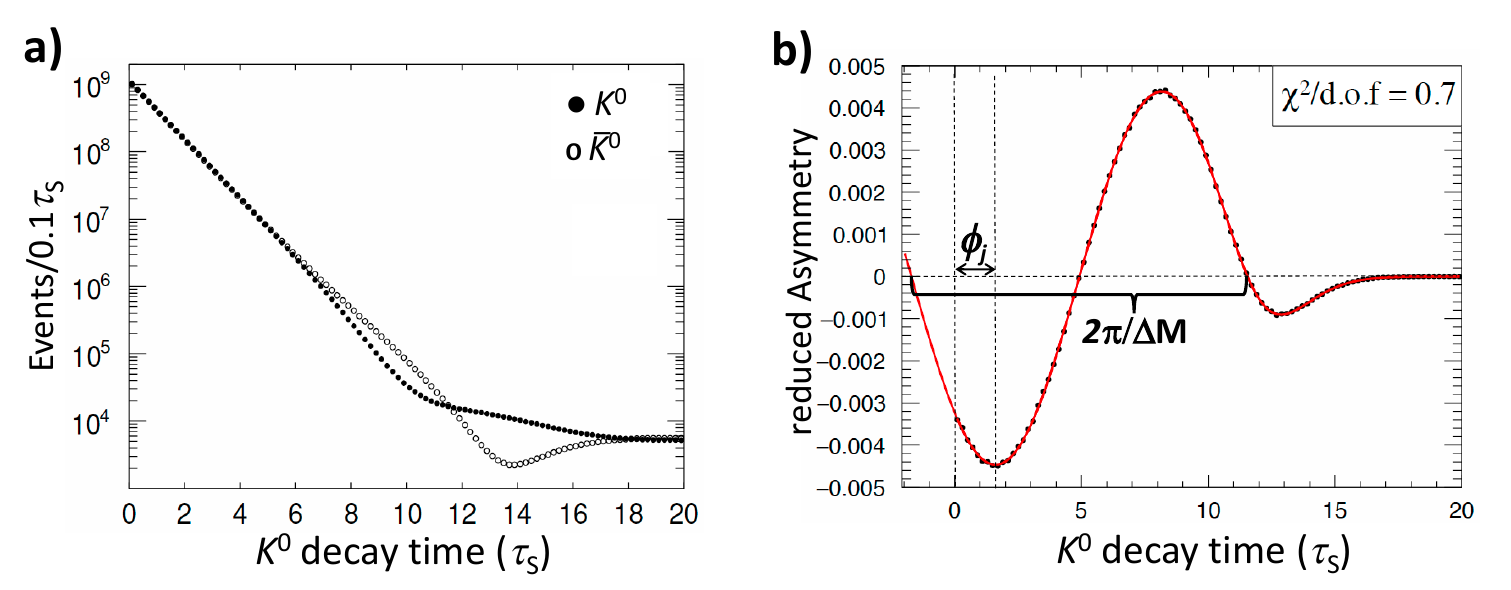}
\caption{\footnotesize {\bf a)} The solid circles show the proper time distribution for simulated strangeness-tagged
  $\Kz(\tau)$$\rt$$\pipi$ decays (the open circles are $\Kzbar(\tau)$$\rt$$\pipi$ decays). 
  {\bf b)} The reduced asymmetry, ${\mathcal A}^{\prime}_{\pipi}$, for the events shown in panel~{\bf a} (from
  ref.~\cite{Zhang:2022lds}).
  }
 \label{fig:A-prime-simul}
\end{figure}

Decay curves for $\Kz(\tau)$$\rt$$\pipi$ and $\Kzbar(\tau)$$\rt$$\pipi$ for simulated BESIII events that were generated
with PDG values of the  $\eta_{+-}$ magnitude and phase~\cite{Zhang:2022lds} are shown in Fig.~\ref{fig:A-prime-simul}a, where there
are strong interference effects between the steeply falling $\KS$$\rt$$\pipi$ and nearly flat $\KL$$\rt$$\pipi$ amplitudes that
have opposite signs for $\Kz(\tau)$- and $\Kzbar(\tau)$-tagged decays. The difference between the interference terms is
displayed in a plot of the reduced asymmetry~\cite{Apostolakis:1999zw} $\mathcal{A}^\prime_{\pi\pi}$,
where\footnote{In the unmodified asymmetry,
  $\mathcal{A}_{\pi\pi}$=$[\bar{N}(\Kzbar(\tau))$$-$$N(\Kz(\tau))]/[\bar{N}(\Kzbar(\tau))$+$N(\Kz(\tau))]$,
  the cos$(\Delta M\tau$$-$$\phi_j)$ oscillation term is multiplied by a factor $e^{\smallonehalf\Delta\Gamma\tau}$
    that, when displayed, emphasizes low-statistics, long-decay-time events that, in fact, have
    little influence on the $\phi_j$ determination.}
\begin{eqnarray}
  \label{eqn:KKbar-asymm-CPT} 
        {\mathcal A}^\prime_{\pi\pi}&\equiv&
        \frac{\bar{N}(\Kzbar(\tau))-N(\Kz(\tau))}{\bar{N}(\Kzbar(\tau))+N(\Kz(\tau))}e^{-\smallonehalf\Delta\Gamma\tau}\\
    \nonumber
    &~&~~~~~~=2\Re\eps_L e^{-\smallonehalf\Delta\Gamma\tau}-2\frac{|\eta_{j}|\cos(\Delta M\tau-\phi_{j})}
              {1+|\eta_{+-}|^2e^{\Delta\Gamma\tau}},
\end{eqnarray}
and is shown for the simulated data in  Fig.~\ref{fig:A-prime-simul}b, where $\phi_j$ shows up as a phase shift in the
interference-generated cosine-like oscillation.

The  simulated data shown in Figs.\,\ref{fig:A-prime-simul}\,a\,\&\,b correspond to 3.8B~tagged $K$$\rt$$\pipi$ decays
that are almost equally split between $\jpsi$$\rt$$\Km\pip\Kz$ and $\jpsi$$\rt$$\Kp\pim\Kzbar$ that were generated 
with $\Delta\phi^{\mathcal{CPT}}$=0\, and $\phi_{\rm SW}$=\,43.5$^\circ$. This corresponds to what one would expect for a
total of $10^{12}$~$\jpsi$ decays in a detector with realistic vertex and momentum  resolutions that covered a
$|\cos\theta|\le 0.85$ solid angle.  The red curve in the figure is the result of a fit to the data that determined~\cite{Zhang:2022lds}
\begin{equation}
  \phi_{\rm SW}+\Delta\phi^{\mathcal{CPT}} = 43.51^\circ\pm 0.05^\circ,
\end{equation}
where the errors are statistical only. According to eqn.\,\ref{eqn:DeltaM_veltaPhi}, this $\mathcal{O}(0.1^\circ)$
precision would correspond to $|M_{\Kzbar}$$-$$M_{\Kz}|$\,$\approx$\,4$\times$10$^{-17}$\,MeV. 

This would be an order of magnitude improvement over the existing eqn.\,\ref{eqn:DeltaM_veltaPhi} limit that is based on a
1995 result from Fermilab experiment E773~that used $\sim$2\,M $K$$\rt$$\pipi$ and $\sim$0.5\,M $K$$\rt$$\piz\piz$ decays produced
downstream of a regenerator located in a high-energy $\KL$ beam~\cite{Schwingenheuer:1995uf}, and a 1999 result from the CPLEAR
experiment at CERN~\cite{Apostolakis:1999zw} that used  $\sim$35\,M-event samples of tagged $\Kz(\tau)$$\rt$$\pipi$ and
$\Kzbar(\tau)$$\rt$$\pipi$ decays produced via the annihilation of stopped antiprotons in a high-pressure gas hydrogen target
into  $K^{\mp}\pi^{\pm}\Kz(\Kzbar)$ final states:
\begin{eqnarray}
  \label{eqn:E773-delCPT}
  {\rm E773:~~}~~~\phi_{\rm SW}+\Delta\phi^{\mathcal{CPT}}&=&42.94^\circ\pm 0.58^\circ {\rm (stat)}\pm 0.49^\circ  {\rm (syst)}\\
   \nonumber
  {\rm CPLEAR:}~~~\phi_{\rm SW}+\Delta\phi^{\mathcal{CPT}}&=&42.91^\circ\pm 0.53^\circ {\rm (stat)}\pm 0.28^\circ  {\rm (syst)}.
\end{eqnarray}

The CPLEAR systematic error includes a $0.19^\circ$ component associated with the effects of regeneration in the
high-pressure hydrogen gas in the stopping target and the walls of its containment vessel. In the E773 measurement,
the phase off-set that was introduced by the regenerator was unavoidable and large, $\phi_{\rm regen}$$\sim$$-$130$^\circ$, and
was evaluated from a dispersion relation with an assigned error of $0.35^\circ$~\cite{Briere:1995tw}.
In an $\ee$ collider operating at the $\jpsi$ with a detector optimized for \C\Par~searches in hyperon decays and \C\Par\T~tests,
the production and a significant fraction of the $K$$\rt$$\pipi$ decays would occur in a vacuum, and the effects of
regeneration on decays that occur outside the vacuum could be measured in specially configured data runs, as was done, {\em e.g.},
by the CPLEAR group~\cite{CPLEAR:1997fwa}).

\subsection{The Bell Steinberger relation}
\label{sec:bell-steinberger}

The above discussion is not entirely rigorous and ignores some corrections\footnote{These are associated with
       ignoring the term involving $\Im\Gamma_{12}$ in eqn.~\ref{eqn:def-eps}.}
that are small compared to the precision of the KTeV and CPLEAR measurements in eqn.~\ref{eqn:E773-delCPT} but will be important for
future measurements. For these, a procedure devised by John Bell and Jack Steinberger that gives an exact expression for $\Im\delta$ in
terms of measurable quantities and is phase-convention independent should be used~\cite{Bell:1965mn}. 

As mentioned above, a general solution to the Schr\"{o}dinger equation is given by
\begin{equation}
  \label{eqn:general-wave-function}
  \psi(\tau)=\alpha_1 e^{-i\lambda_S\tau}\ket{\KS} +\alpha_2 e^{-i\lambda_L\tau}\ket{\KL},
\end{equation}
where $\alpha_1$ and $\alpha_2$ can be any complex constants subject only to the normalization
condition $|\alpha_1|^2$+$|\alpha_2|^2$=1. The time-dependent probability associated with this wave function is 
\begin{equation}
  |\psi(\tau)|^2=|\alpha_1|^2e^{-\Gamma_S\tau}+|\alpha_2|^2e^{-\Gamma_L\tau}
  +2 \Re\big(\alpha^*_1\alpha_2 e^{-{\scriptstyle \frac{1}{2}}(\Gamma_S+\Gamma_L+2i\Delta M)\tau}\braket{\KS}{\KL}\big),
\end{equation}
and the negative of its derivative at $\tau$\,=0 is
\begin{equation}
  \label{eqn:bell-steinberger-lhs}
  -\frac{d|\psi(\tau)|^2}{d\tau}\bigg|_{\tau=0}=|\alpha_1|^2\Gamma_S + |\alpha_2|^2\Gamma_L
  +\Re\big(\alpha^*_1\alpha_2(\Gamma_S+\Gamma_L+2i\Delta M)\braket{\KS}{\KL}\big).
\end{equation}

In the Weisskopf-Wigner formalism, the Schr\"{o}dinger equation is not hermitian and the solutions
do not conserve probability. Instead, the overall normalization has a time dependence given by
\begin{equation}
|\psi(\tau)|^2=|\psi(0)|^2e^{-\Gamma_{\rm tot}\tau};
  ~~\frac{d|\psi(\tau)|^2}{d\tau}=-\Gamma_{\rm tot}e^{-\Gamma_{\rm tot}\tau},
\end{equation}

\noindent
and unitarity requires that
\begin{equation}
 -\frac{d|\psi(\tau)|^2}{d\tau}\bigg|_{\tau=0}=\Gamma_{\rm tot}=\sum_j\Gamma_j=\sum_j |\braket{f_j}{\psi(0)}|^2,
\end{equation}

\noindent
where the summation index $j$ runs over all of the accessible final states $\ket{f_j}$.
Applying this unitarity condition to the eqn.\,\ref{eqn:general-wave-function} wave function gives an independent
expression for the derivative of $|\psi(\tau)|^2$ at $\tau$\,=0:
\begin{eqnarray}
    \label{eqn:bell-steinberger-rhs}
  -\frac{d|\psi(\tau)|^2}{d\tau}\bigg|_{\tau=0}&=&\sum_j|\alpha_1|^2|\braket{f_j}{\KS}|^2 + |\alpha_2|^2\braket{f_j}{\KL}|^2\\
    \nonumber
        &&~~~~~~~~+2\Re\big(\alpha^*_1\alpha_2\braket{f_j}{\KS}^*\braket{f_j}{\KL}\big).
 \end{eqnarray}

\noindent
Equations~\ref{eqn:bell-steinberger-lhs} and~\ref{eqn:bell-steinberger-rhs} have to apply for any values of
$\alpha_1$ and $\alpha_2$ (subject to $|\alpha_1|^2+|\alpha_2|^2=1$), and, thus, the terms multiplying $\alpha^*_1\alpha_2$ in
each expression have to be equal
\begin{equation}
  (\Gamma_S+\Gamma_L+2i\Delta M)\braket{\KS}{\KL} = 2\sum_j\braket{f_j}{\KS}^*\braket{f_j}{\KL},
\end{equation}

\noindent
that, together with eqn.\,\ref{eqn:KLKS-CPT}, becomes

\begin{equation}
  \label{eqn:bell-steinberger}
        {\textstyle \frac{1}{2}}\braket{\KS}{\KL}=
        \frac{\Re\eps}{1+|\eps|^2|} -i\Im\delta
        =\frac{\sum_j\braket{f_j}{\KS}^*\braket{f_j}{\KL}}{ \Gamma_S+\Gamma_L+2i\Delta M },     
\end{equation}

\noindent
which is the {\it Bell-Steinberger relation}.  Here the second-order term in $|\eps|$ is retained because 
with it, the equation is exact. A nonzero value of $\Im\delta$ could only occur if \C\Par\T~ is violated, or
if unitarity is invalid, or if the $M_{ij}$ and/or $\Gamma_{ij}$ elements of the Hamiltonian were time dependent.

Because of the small number of $\KS$ decay modes, the sum on the right-hand side of this equation only contains a few terms and is
manageable. In contrast, the neutral $D$- and $B$-mesons have hundreds of established decay modes and probably a similar number of
modes that have yet to be detected, and any attempt to apply a Bell-Steinberger-like unitarity constraint would be a hopeless task. 

A Bell-Steinberger-relation evaluation of  $\Im \delta$ and $\Re\eps$ by Giancarlo D'Ambrosio and Gino Isidori in
collaboration with the KLOE group that used of all the latest relevant measurements was originally presented in
2006~\cite{DAmbrosio:2006hes} and recently updated in the PDG2020 review~\cite{DAmbrosio:2022pth}. Their results are:

\begin{center}
  \begin{tabular}{lc}
  \hline
  channel           &  $\braket{f_j}{\KS}^*\braket{f_j}{\KL}$~($10^{-5}\Gamma_S$)\\
  \hline
  $\pipi(\gamma)$   &    $112.1\pm 1.0~+~i(106.1\pm 1.0)$  \\
  $\piz\piz$        &     $49.3\pm 0.5~+~i(47.1\pm 0.5)$   \\
  $\pipi\gamma_{E1}$ &    $<0.01$\\
  $\pipi\piz$       &    $0.0\pm 0.2~+~i(0.0\pm 0.2)$  \\
  $\piz\piz\piz$    &    $<0.15$   \\
  $\pi^\pm\ell^\mp\nu$ &  $-0.1\pm 0.2~+~i(-0.1\pm 0.5)$,\\
  \hline
  \end{tabular}
\end{center}
\noindent
where: the $\pi\pi(\gamma)$ values come from the PDG~2020 averages of many  $\eta_{+-}$ and $\eta_{00}$ measurements;
the $\pipi\gamma_{E1}$ limit comes from theory~~\cite{Cheng:1993xd,DAmbrosio:1996lam}; the $\pi\pi\pi$  results are
based on thirty-year-old $\BR(\KS\rt\pipi\piz)$ measurements~\cite{Zou:1996ks,Adler:1997pp} and a
2013 KLOE upper bound on $\BR(\KS\rt\piz\piz\piz)$~\cite{Babusci:2013tr};  and the $\pi^\pm\ell^\mp\nu$ values come from
twenty-five-year-old limits on $\DelS$\,=$-\Delta Q$ processes from CPLEAR~\cite{CPLEAR:1998ohr}.  

\begin{figure}[!bp]
\begin{center}
     \includegraphics[height=0.4\textwidth,width=0.8\textwidth]{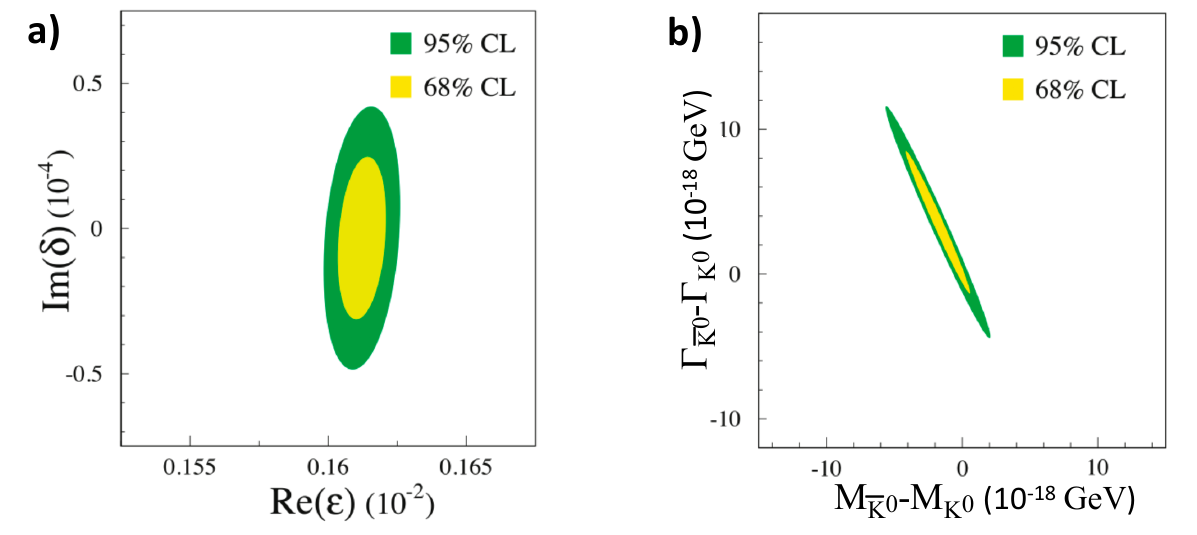}
     \caption{\footnotesize  {\bf a)} The 68\% and 95\% confidence level allowed region for
       $\Im\delta$ and $\Re\eps$ from a Bell-Steinberger analysis.
       {\bf b)} The corresponding allowed regions in $\Delta\Gamma$=\,$\Gamma_{\Kzbar}$$-$$\Gamma_{\Kz}$ and
       $\Delta M= M_{\Kzbar}-M_{\Kz}$. From ref.~\cite{Zyla:2020zbs}.
     }
     \label{fig:bell-steinberger}
\end{center}
\end{figure}

The bottom line of this analysis is
 \begin{equation}
   \label{eqn:BS-results}
   \Im\delta = (-0.3\pm 1.4)\times 10^{-5}~~~\Rightarrow~~~|M_{\Kzbar}-M_{\Kz}|<4\times 10^{-16}\,{\rm MeV},
 \end{equation}
 and the $\Im\delta$\,{\it vs.}\,$\Re\eps$ and $\Gamma_{\Kzbar}$$-$$\Gamma_{\Kz}$\,{\it vs.}\,$M_{\Kzbar}$$-$$M_{\Kz}$
 allowed regions are shown in Figs.\,\ref{fig:bell-steinberger}\,a\,\&\,b, respectively.
 The limit on $|M_{\Kzbar}$$-$$M_{\Kz}|$ is nearly the same as the eqn.\,\ref{eqn:DeltaM_veltaPhi} limit that comes from simply
 comparing the PDG~2020 world average values for $\phi_{+-}$ and $\phi_{\rm SW}$.  This is not surprising since in both
 analyses the  $\pi\pi$ channels dominate and the sensitivities are limited by the same $\eta_{+-}$ and $\eta_{00}$ measurement
 errors. Moreover the contributions from the three-body modes and their errors are smaller than those from the
 $\pi\pi$ modes.   However, when compared to the projected errors of measurements with 10$^{12}$~ $\jpsi$ events,
 the errors on these auxiliary contributions are not small, and, thus, future progress in  Bell-Steinberger-relation
 analyses will require that improvements in the $\eta_{+-}$ and $\eta_{00}$ phase measurements are accompanied by
 similar improvements in the precision of the  other terms.

The $\pipi\piz$ entry in the above table are from statistics-limited time-dependent Dalitz-plot analyses of
$\KS$-$\KL$ interference effects in strangeness-tagged $\Kz(\tau)$ and $\Kzbar(\tau)$ decays to $\pipi\piz$.
Similarly, the $\pi\ell\nu$ entry is limited by the measured values lepton charge asymmetries that were used to establish
limits on violations of the $\DelS$\,=\,$\Delta Q$ rule that also used strangeness-tagged kaons. The precision of
both of these entries could be improved with the higher statistics data samples that would be available at a future facility.

That leaves the $\piz\piz\piz$ entry, which is not a limiting factor now, but will be if (or when) the $\pipi$ and $\piz\piz$
and other entries are improved by an order of magnitude. It is based on a $\BR(\KS$$\rt$3$\piz)$$<$2.6$\times$10$^{-8}$ upper
limit.\footnote{The SM expectation is $\sim$10$\times$\,lower at
$\BR^{\rm SM}(\KS$$\rt$$3\piz)$$\approx$$|\eps|^2(\tau_S/\tau_L)\BR (\KL$$\rt$$3\piz)$\,=1.7$\times$10$^{-9}$.}
from the KLOE experiment that used tagged $\KS$ mesons in a 1.7\,billion $\ee$$\rt$$\phi$$\rt$$\KS\KL$ event data sample
produced at the DAFNE $\phi$ factory~\cite{Babusci:2013tr}. The KLOE analysis detected 75\,M $\KS$$\rt$$\piz\piz$ decays but
found zero $\KS$$\rt$3$\piz$ events
In the absence of any prospects for a next-generation $\phi$ factory, further improvements of this entry will not be easy.

\section{The strange history of the Standard Model}

In the 1950s and 60s, measurements of the ``strange'' properties of of the newly discovered hyperons and kaons taught us almost
all of what was needed to develop the Standard Model, including:
\begin{description}
\item{\em Flavor Quantum Numbers}:~``Strangeness'' was the first of the SM's ``flavors'' to be identified;
\item{\em Particle-antiparticle Mixing}:~$K^0$$\leftrightarrow$$\bar{K}^0$-like transitions also occur with charmed and
  bottom particles, and related processes occur with neutrinos;
\item{\em Parity Violation}:~left-right symmetry violating processes were first seen in  $K$-meson decays
  and are now a defining characteristic of the Weak Interactions;
\item{{$\mathcal{CP}$}\em ~Violation}: the first sign of matter and antimatter differences was the observation of
  $\KL$$\rt$$\pipi$ decays;
\item{\em Flavor mixing}: the suppression of the $\Lambda$$\rt$$p e^-\bar{\nu}$ decay rate below that inferred from
 the half-life for $n$$\rt$$p e^-\bar{\nu}$ decay was the first sign of flavor-mixing, and
 the precursor of the CKM quark-mixing and PMNS neutrino-mixing matrices;
\item{\em Quark Model}: the Isospin-Strangeness patterns in the baryon  octet~\&~decuplet and the meson octets
  led to the discovery of $SU(3)$ and fractionally charged quarks;
\item{\em Color}: the existence of $\Omega^-$ baryon that, in the quark model, was comprised of three identical $s$-quarks
     in a symmetric state, indicated the existence of a hidden ``color'' quantum number;
\item{\em GIM Mechanism}: the strong suppression of strangeness-changing neutral-current kaon decays led to prediction
  of the  existence of the $c$-quark;
\item{\em Six quarks}: the accommodation of \C\Par~violation into quark-flavor-mixing matrix required a minimum of six quark flavors.
\end{description}
This all occurred before the 1974 discovery of the $\jpsi$.   The only subsequent discovery that could not be accounted for
by the Standard Model was neutrino mixing, which was predicted by Bruno Pontecorvo~\cite{Pontecorvo:1957qd} in 1958, who was
inspired by the 1955 discovery of $K$-$\bar{K}$ mixing.\\

After the $c$-quark discovery, interest in strangeness physics abated and, after the 1990s, none the world's major
particle physics developed the capability of producing enough kaons to match, much less improve on, the E773 and CPLEAR results,
choosing instead to focus on experiments with huge numbers of charmed and beauty particles that can that address many interesting
subjects, but will never be able to test thes \C\Par\T~ theorem with the exquisite sensitivity that is uniquely provided by
the neutral $K$-meson system  or search for new sources of \C\Par~violation in hyperon decay. As noted in this report, a facility
that could produce $\sim$10$^{12}$ $\jpsi$ per year (or so) would provide opportunities to make order of magnitude sensitivity improvements
over earlier experiments in both areas, all with relatively modest costs, especially for  a laboratory with existing intense
electron and and positron sources.

Strange particles taught us a lot in the past, maybe they'll teach us even more in the future.

\section{Acknowledgement}
The author thanks the organizers and staff of the 50 Years Discovery of the $J$ Particle Symposium for inviting me to give this
talk and their kind hospitality during our visit to Beijing. I also  acknowledge numerous informative discussions about the
material contained in this report with my BESIII colleagues Andrzej Kupsc, Haibo Li, Jian-Yu Zhang and Xiaorong Zhou.
This work was supported in part by the National Research Foundation of
Korea under Contract No.~NRF-2022R1A2C1092335, and the PIFI program of the Chinese Academy of Science.

\appendix

\section{Appendix on monochromatic beams}

\begin{figure}[h!]
\centering
\includegraphics[width=0.99\textwidth]{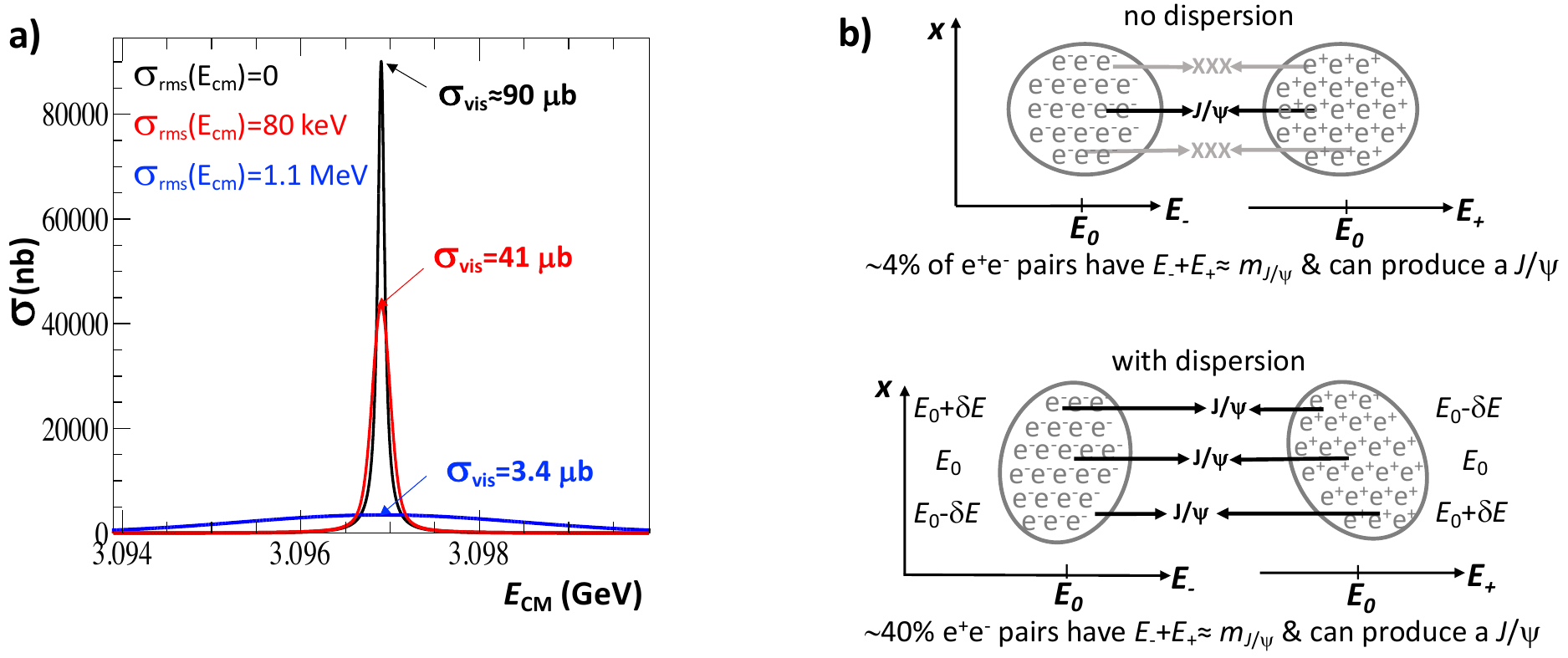}
\caption{\footnotesize {\bf a)} The visible  $\ee$$\rt$$\jpsi$ lineshape for different values of the $\Ecm$ resolution
   (provided by Xiaoshuai Qin).
  {\bf b)} Usually colliders operate with zero dispersion at the interaction point, with no correlation between
  a beam particle's energy and its horizontal position, as shown in the upper part of the figure. In
  monochromator operation, dispersion is introduced as shown in the lower section, so that lower energy particles
  in one beam preferentially collide with higher energy particles in the other beam (and {\it vice versa}).
 }
 \label{fig:monochromator}
 \end{figure}

A possible method to increase the $\jpsi$ event rate without increasing beam currents and is specific to a $\ee$$\rt$$\jpsi$ c.m.~collider,
would involve decreasing the effective $\ee$ c.m.~energy spread.  In circular $\ee$ storage rings, statistical fluctuations in the number
of synchrotron X-rays that are radiated turn-to-turn create an irreducible energy spread in the circulating $e^+$~and~$e^-$ beams
that, in the BEPCII collider, is  $\sigma_{\rm rms}(E_{\rm beam})$\,$\approx$\,5$\times$10$^{-4}E_{\rm beam}$. Thus, although the $\jpsi$
is a Breit Wigner resonance with a peak cross-section of 90~$\mu$b and a FWHM width\footnote{The corresponding
  rms~width is $\sigma_{\rm rms}$\,$\approx$40~keV.}
of $\Gamma$=\,93\,keV, as shown as a black curve in Fig.~\ref{fig:monochromator}a, the c.m.~energy spread in BESIII
dilutes this narrow peak to a visible line-shape that is a Gaussian with a $\sigma_{\rm vis}(\ee$$\rt$$\jpsi)$\,=\,3.4$\mu$b peak
cross-section with a 1.1\,MeV rms energy spread, as shown
by the blue curve in Fig.~\ref{fig:monochromator}a.  When the collider operates at the $\jpsi$ mass peak, about
96\% of the $\ee$ combinations have an energy sum that is either too low or too high to produce a $\jpsi$, as
illustrated by the cartoon in the upper panel of Fig.~\ref{fig:monochromator}b.

A monochromator scheme that was first proposed by Alfredo Renieri~\cite{Renieri:1975wt} and, independently, by
Igor~Protopopov, Alexander Skrinsky~\&~Alexander Zholents~\cite{Protopopov:1979tn}, could improve the $\Ecm$
resolution with a commensurate increase in the visible $\jpsi$ peak cross section. This would be accomplished by
introducing some energy dispersion into both beams at their collision point, {\it i.e.}, a horizontal position
dependence of the beam-particle energy, as shown in the lower portion of Fig.~\ref{fig:monochromator}b. Here the
dispersion is arranged so the low-energy side of the $e^-$ beam collides with the high energy side of the $e^+$
beam profile, and {\it vice versa}, thereby reducing the effective c.m. energy spread. If $\sigma_{\rm rms}(\Ecm)$
can be reduced to~$\approx$\,80~keV, the peak cross section would be ten times higher at
$\sigma_{\rm vis}(\ee$$\rt$$\jpsi)$$\approx$\,41$\mu$b as shown as a red curve in Fig.~\ref{fig:monochromator}a. A lattice
design by Zholents provided a proof of principle that an order of magnitude improvement in the $\Ecm$ resolution
could be achieved in a head-on collider~\cite{Zholents:1992ua}.

A variation of this technique to large-crossing-angle colliders that includes angular dispersion in addition to
energy dispersion was proposed by Valery Telnov~\cite{Telnov:2020rxp}. He concludes that an $\ee$ collider
operating with a very large crossing angle, $\sim$300\,mr---four times SuperKEKB's 83\,mr crossing angle---could
reduce the effective mass resolution by as much as two orders of magnitude, in which case the
visible cross section at the $\jpsi$ peak would be the full 90\,$\mu$b, with no apparent luminosity penalty.
Although such a large crossing angle would be a major complication for a general purpose facility, where some
physics goals might require hermetic $4\pi$ detector coverage and/or meticulous micro-vertex resolution, it would
probably be tolerable for most $\jpsi$-related topics, including the above-described \C\Par\T~test.

\bibliographystyle{unsrt}
\bibliography{Jpsi-50_solsen} 
\end{document}